\documentclass[12pt,preprint]{aastex}

\begin{document}
\title{Galaxy Spin Statistics and Spin-Density Correlation}
\author{Jounghun Lee}
\affil{Institute of Astronomy and Astrophysics, Academia Sinica,
Taipei, Taiwan}
\email{taiji@asiaa.sinica.edu.tw}
\author{Ue-Li Pen}
\affil{Canadian Institute for Theoretical Astrophysics,
Toronto, Ont. M5S 3H8, Canada}
\email{pen@cita.utoronto.ca}

\newcommand{\etal}{{\it et al.}}
\newcommand{\beq}{\begin{equation}}
\newcommand{\eeq}{\end{equation}}
\newcommand{\ben}{\begin{eqnarray}}
\newcommand{\een}{\end{eqnarray}}
\newcommand{\hbL}{\hat{\bf L}}
\newcommand{\hbrL}{\hat{\bf L}^{\gamma}}
\newcommand{\hrL}{\hat{L}^{\gamma}}
\newcommand{\hbS}{\hat{\bf S}}
\newcommand{\bS}{{\bf S}}
\newcommand{\bL}{{\bf L}}
\newcommand{\be}{{\bf e}}
\newcommand{\hL}{\hat{L}}
\newcommand{\hT}{\hat{T}}
\newcommand{\hk}{\hat{k}}
\newcommand{\hr}{\hat{r}}
\newcommand{\hQ}{\hat{Q}}
\newcommand{\bk}{{\bf k}}
\newcommand{\bA}{{\bf A}}
\newcommand{\bu}{{\bf u}}
\newcommand{\by}{{\bf y}}
\newcommand{\bx}{{\bf x}}
\newcommand{\bt}{{\bf t}}
\newcommand{\hby}{\hat{\bf y}}
\newcommand{\hbd}{\hat{\bf d}}
\newcommand{\hbx}{\hat{\bf x}}
\newcommand{\hbr}{\hat{\bf r}}
\newcommand{\hbt}{\hat{\bf t}}
\newcommand{\br}{{\bf r}}
\newcommand{\bfv}{{\bf v}}
\newcommand{\bq}{{\bf q}}
\newcommand{\bQ}{{\bf Q}}
\newcommand{\bV}{{\bf V}}
\newcommand{\bU}{{\bf U}}
\newcommand{\bI}{{\bf I}}
\newcommand{\bT}{{\bf T}}
\newcommand{\tT}{\tilde{T}}
\newcommand{\txi}{\tilde{\xi}}
\newcommand{\tA}{\tilde{A}}
\newcommand{\tC}{\tilde{C}}
\newcommand{\tbT}{\tilde{\bf T}}
\newcommand{\tbaT}{\tilde{\bf T}^{\alpha}}
\newcommand{\tbbT}{\tilde{\bf T}^{\beta}}
\newcommand{\tbrT}{\tilde{\bf T}^{\gamma}}
\newcommand{\taT}{\tilde{T}^{\alpha}}
\newcommand{\trT}{\tilde{T}^{\gamma}}
\newcommand{\tbeT}{\tilde{T}^{\beta}}
\newcommand{\tbC}{\tilde{\bf C}}
\newcommand{\hbT}{\hat{\bf T}}
\newcommand{\hbQ}{\hat{\bf Q}}
\newcommand{\p}{\phi}
\newcommand{\hS}{\hat{S}}
\newcommand{\pr}{\prime}
\newcommand{\lam}{\lambda}
\newcommand{\hlam}{\hat{\lambda}}

\begin{abstract}

We present a theoretical study of galaxy spin correlation statistics,
with detailed technical derivations.  We also find an expression for 
the spin-density cross-correlation, and apply that to the Tully galaxy 
catalog.  The observational results appear qualitatively consistent with 
the theoretical predictions, yet the error bars are still large.
However, we expect that currently ongoing large surveys such as the 
Sloan Digital Sky survey (SDSS)  will enable us to make a precision 
measurement of these correlation statistics in the near future.  
These intrinsic galaxy alignments are expected to dominate over the 
weak lensing signal in SDSS, and we present the detailed algorithms 
for the density reconstruction for this case.

These observables are tracers of the galaxy-gravity 
interaction, which may provide us deeper insights into the 
galaxy formation and large scale matter distribution as well. 
\end{abstract} 
\keywords{galaxies:statistics --- large-scale structure of universe}

\section{INTRODUCTION}

The origin and evolution of the galaxy angular momentum, i.e., the
galaxy spin has been the subject of many studies in the last 
century.  Hoyle (1949) suggested an original idea that the origin 
of the rotational galaxy motion could be ascribed to the 
gravitational coupling with the surrounding galaxies. 
Sciama (1955) applied Hoyle's idea to his theory for the 
formation of galaxies in a steady-state universe model. 

It was Peebles (1969) who first quantitatively examined Hoyle's idea 
in the gravitational instability picture. He argued that   
the shear effect due to the primordial tidal torquing 
from the neighbor matter distribution should be mainly responsible
for the acquisition of the angular momentum by a proto-galaxy.  
He pointed out that the alternative models 
for the origin of the galaxy angular momentum such as the 
initial vorticity model and the primeval turbulence model 
proposed by von Wiezsacker (1951) and Gamow (1952) to 
a wrong prediction of too early formation of galaxies. 
Assuming a spherical symmetry of a proto-galaxy, he analyzed quantitatively 
the growth rate of the magnitude of the galaxy angular momentum in the frame
of the linear perturbation theory, and drew a conclusion that the galaxy 
angular momentum grows as proportional to the second order perturbation 
($\propto t^{5/3}$ for a $\Omega = 1$ universe).  

In contrast,  White (1984) showed that the proto-galaxy angular momentum 
grows at first order ($\propto t$ for a $\Omega = 1$ universe) unless 
the restrictive condition of the spherical symmetry is imposed on 
proto-galactic sites, which was originally contended by Doroshkevich (1970). 
He expanded Doroshkevich's contention in detail by means of 
the linear perturbation theory described by the Zel'dovich approximation, 
and confirmed that the proto-galactic angular momentum is generated by 
the misalignment between the proto-galactic inertia tensor and the local 
gravitational shear tensor, and grows to first order during 
the linear phase. He confirmed his results by N-body simulations.   

Heavens \& Peacock (1988) analyzed the correlation of the galaxy  
angular momentum with  the local density maxima in the linear 
regime, and concluded that the total angular momentum in the linear 
regime is almost independent of the height of the density peaks 
(see also Hoffman 1986, 1988).  Catelan \& Theuns (1996) extended 
the Heavens-Peacock works and calculated the expectation value of the 
angular momentum assuming an ellipsoidal proto-galaxy centered on a 
peak of the Gaussian density field using White's formula.  
 
While all these studies concentrated on the magnitude of the angular
momentum, the total angular momentum, or even the fraction of virial
energy in rotation, is very difficult to observe.  On the other hand, 
the direction of the angular momentum, i.e., the galaxy 
spin axis can be measured only from the position angle on 
the sky and the projected axis ratio, which can be implemented 
for very large surveys.  
Therefore, the galaxy spin axis could provide more useful statistics 
that can be easily tested against real observational data. 

Recently Lee \& Pen (2000, hereafter LP00) pointed out that the 1st-order 
linear perturbation theory predicts preferential alignments of the 
galaxy spin axis along with the  2nd principal axis of the local 
gravitational shear tensor, and suggested a unique statistical model
that uses the galaxy spin axis as a tool to reconstruct the initial 
density field.  Their theory is based on two basic assumptions: 
First, the spin axis of a galaxy aligns well with that of the 
underlying dark halo. Second, the galaxy spin aligns with the 2nd 
principal axis of the local gravitational shear tensor 
to a detectable degree. 
 
The first assumption is generally accepted as a reasonable working 
hypothesis for the spiral galaxies. A spiral galaxy is a highly flattened 
disk with its plane perpendicular to the direction of the 
underlying halo spin axis in most galaxy-formation theories  
(e.g., Mo, Mao, \& White 1998).  
There is also an observational clue to this assumption.   
The galaxy spin can be observed to much larger radii than the 
galaxy radius through radio emission of the gas, and the spin 
direction of a spiral galaxy has been seen to change 
only very modestly as one moves to larger radii. This suggests 
that the spin axis of a galaxy is well correlated with 
that of the whole halo.
 
Meanwhile the second assumption should work subject to numerical testing. 
In fact several N-body experiments have already shown that the dark 
halos have preferred direction in the spin orientation, and that this 
preferential spin alignment has likely a primordial origin.
Dubinski (1992) for the first time found this preferential spin 
alignment in N-body simulations of dark halos. By comparing simulations 
in the absence and presence of a cosmological tidal field, he investigated 
the effect of the initial shear on the spin orientation. He concluded 
that the numerically detected preferential spin alignment measured in 
dark halos resulted from the shear effect due to the linear regime 
tidal torque. LP00 directly calculated the correlation in direction  
between the Lagrangian and the Eulerian angular momentum of dark halos 
measured in N-body simulations, and confirmed that the  
linear theory prediction for the orientation of the halo angular 
momentum is quite a good approximation.    
    
The central concept of LP00 is that one can use this preferential 
spin alignment, if it really exists, as a linking bridge between 
the initial matter distribution of the universe and the observable 
unit galaxy-spin field.  They have provided a mathematical algorithm 
to reconstruct the initial shear and density fields from the
observable galaxy spin axes.  Conventionally the peculiar velocity 
or the weak lensing shear fields have been used to reconstruct 
the total mass density field. This new method for the density reconstruction 
using the galaxy spin field is believed to be advantageous for a 
couple of reasons: First, the galaxy spin axis is relatively easier 
to measure observationally. Second, it is free of the standard 
galaxy biasing. Third, it allows the reconstruction of the full 
three dimensional density field.
 
Very recently, numerical study of the galaxy spin or galaxy ellipticity 
alignment due to the local gravitational shear has become quite topical.   
The flurry of recent activities  (Croft \& Metzler 2000; 
Heavens, Refregier, \& Heymans 2000; 
Catelan, Kamionkowski, \& Blandford 2000; 
Crittenden \etal\ 2000a; Crittenden \etal\ 2000b)  
is motivated partly by the statistical search in blank fields for 
weak lensing signal, for which the intrinsic galaxy alignment plays   
a role of systematic error. The gravitational shear effect  
on the galaxy spin axis due to the initial tidal torquing 
is local, distinguished from the weak lensing 
shear which can change only the apparent orientation 
of the galaxy spin axis due to the distant intervening matter far 
from both the source and the observer. 
From here on, the intrinsic cosmic shear is referred to as the 
local gravitational shear. 

Heavens et al. (2000) calculated the intrinsic ellipticity 
correlation of dark halos found in high-resolution N-body 
simulations. They implied that at small redshift the  intrinsic 
ellipticity correlation due to the local cosmic shear effect 
dominates the  correlation signal due to the weak 
lensing effect. Crittenden \etal\ (2000a) reanalyzed the results 
of Heavens \etal\ and demonstrated that the results of Heavens \etal\ 
from high-resolution N-body simulations in fact indicate stronger 
intrinsic spin alignments than that of LP00 from low-resolution 
simulations.  Croft \& Metzler (2000) also detected the intrinsic 
correlation of projected ellipticities of dark halos in high-resolution 
N-body simulations, and also showed that the correlation signal is 
not strongly affected by the resolution of the simulations. Actually  
they found by comparing two simulations of different resolutions 
that the simulations of higher resolution found more intrinsic correlations.  

Observationally, the question of galaxy spin alignment has received
periodic attentions.
The history of observational search for galaxy alignment traces 
back to the 19th century, and has been marked by checkered records 
\citep{str-str78,gre-etal81,bin82,hel-sal82,hel84,dek85,lam-etal88,
fli88,hof-etal89,kas-oka92,mur-lam92,god94,han-etal95,cab-dic99}.  
For a review of the history of the field, 
see Djorgovski (1987) and Cabanela \& Aldering (1998). 
However, past observational searches for galaxy alignment suffered 
from the small sample sizes \cite{cab-dic99}. It was only very recently 
that positive and reliable signals of galaxy alignments have been 
detected from large galaxy samples.  
Pen, Lee, \& Seljak (2000,  hereafter PLS00) have reported a 
tentative detection of the intrinsic spin correlation signal 
from the Tully galaxy catalog.  The observed signal turns out to be 
significant at the $97\%$ confidence level with the amplitude 
of order of $1\%$ at $1h^{-1}$ Mpc, which is consistent with the 
theoretical predictions made by PLS00. 
Brown et al. (2000) also detected the intrinsic alignment 
in galaxy ellipticities using the SuperCOSMOS Sky survey data.  
They showed that their results agree well with the linear theory  
predictions on the galaxy preferential alignment 
(Crittenden \etal\ 2000a; LP00).  
The observed and simulated amplitudes of correlations are expected 
to be stronger than the weak lensing effects for surveys such as 
Sloan Digital Sky survey (SDSS), and thus a quantitative analysis 
of intrinsic alignments must be completed before one can
attempt to measure weak lensing shears within SDSS.
These positive observational results hint at a possible detection 
of the spin-density cross correlation signal which will be 
addressed here. 

The theory proposed by LP00 that the linear shear and density fields 
can be reconstructed using the detectable intrinsic galaxy spin 
alignments is quite speculative, based on many simplifying assumptions. 
The idea of LP00 must go through thorough observational and numerical 
testings in the future.   However, the recent observational detections of 
intrinsic galaxy alignment and the agreement of the strength of the observed 
signals with the theoretical predictions encourage us to have a prospect 
for the plausibility of our theory and its application to the real 
universe.  If, as predicted, the intrinsic alignment signal indeed 
dominates the weak lensing signal in shallow surveys like SDSS, 
the extraction of intrinsic shear becomes more plausible than that 
of the weak lensing.

In this paper, we present the galaxy spin correlation statistics with 
technical details.  In $\S 2$, we review the mathematical algorithms  
for the density reconstruction given by LP00 in greater detail for 
the reader's thorough understanding of our previous and future works.  
In $\S 3$ we review the spin-spin correlation statistics, and provide 
an analogous spin-density correlation statistics. 
In $\S 4$ we compare the theoretical estimates given in $\S 3$ 
with the observed signals. 
In $\S 5$ the results are summarized and final conclusions are drawn. 
We relegate the detailed calculations and derivations to Appendices A - J.

\section{DENSITY RECONSTRUCTION}

In the standard gravitational instability picture, a proto-galaxy 
acquires its angular momentum from the local gravitational shears 
due to the tidal interaction with the surrounding matter.  
The angular momentum of this proto-galaxy gradually evolves till the 
proto-galactic region reaches the moment of recollapse.  
On recollapse, separated out from the rest of the universe, 
its angular momentum would be approximately conserved afterwards. 
In other words, the galaxy angular momentum is expected to preserves 
its initial dependence on the local shear tensor fairly well 
that has been acquired during the linear regime.  
It is worth mentioning that the galaxy merging or 
secondary infall does not break the dependence of the galaxy angular 
momentum on the initial shears since the total rotational angular 
momentum after merging or infall process is the result of the 
constituent orbital angular momentum of the galaxies combined,  
which depends on the initial shear tensors.  Similarly, the impact
parameter of a collision is also determined by the shear field.  Thus,
what changes by those processes is only the smoothing scale of the 
intrinsic shear which correlates with the galaxy angular momentum.   

It is true that one cannot expect the linear theory to   
fully describe the evolution of the galaxy angular momentum. 
Nonlinear effects such as {\it galaxy-galaxy interaction} 
and etc. may modify the galaxy angular momentum during the 
subsequent evolutionary stages. Nevertheless recent numerical 
simulations have found that in fact the linear theory predictions 
for the direction of galaxy spins are in fairly good agreement 
with numerical results (Dubinski 1992; LP00).  Thus, we base our 
study of the direction of the galaxy angular momentum on  
the linear perturbation theory. 

White (1984) and Catelan \& Theuns (1996) have shown that in the 1st 
order linear perturbation theory described by the Zel'dovich approximation, 
the galaxy angular momentum in Lagrangian space is expressed as 
\beq
L_i(t) = - S^{2}(t)\frac{dD(t)}{dt}\epsilon_{ijk} T_{jl}I_{lk},
\label{eqn:1stl} 
\eeq
where $S(t)$ is the expansion factor, $D(t)$ describes 
the growing mode of the density perturbations, 
$\bI =(I_{lk})=(\int q_l q_k d^{3}\bq )$ is the inertia tensor 
of a proto-galactic site in Lagrangian space,  
$\bT =(T_{jl})=(\partial_j\partial_l\phi)$ is the local shear 
tensor defined as the second derivative of the gravitational potential, 
$\phi$ smoothed on a galactic scale of $R$.  

Rotating the frame into the principal axis of the local shear 
tensor, $\bT$,  we can reexpress equation (\ref{eqn:1stl}) in terms 
of the three eigenvalues, $\lambda_1, \lambda_2, \lambda_3$ of $\bT$  
such that
\beq
L_1 \propto (\lambda_2 - \lambda_3 )I_{23},\hspace{0.5cm} 
L_2 \propto (\lambda_1 - \lambda_3 )I_{31},\hspace{0.5cm}
L_3 \propto (\lambda_1 - \lambda_2 )I_{12},\hspace{0.5cm} 
\label{eqn:eigl} 
\eeq
where the three eigenvalues are ordered to be   
$\lambda_1 > \lambda_2 > \lambda_3$. 
We note three important implications of equation (\ref{eqn:eigl}). 
First, if a proto-galactic region is spherically symmetric 
(corresponding to $I_{12} = I_{23} = I_{31} = 0$),  
then the region gains no angular momentum at 1st order.  
Second, if the principal axis of the inertia tensor, $\bI$, is 
aligned perfectly with that of the shear tensor, $\bT$, then 
the off-diagonal elements of the inertia tensor is zero in the 
shear principal axis frame, 
no angular momentum generated at 1st order, either.   
Third,  if the proto-galactic region is non spherical, and the 
principal axis of the inertial tensor is misaligned with that 
of the shear tensor, then one can expect the region to acquire 
a net angular momentum vector with $L_2$ being dominant since
$\lambda_1-\lambda_3$ is always bigger than the other two differences. 
In other words,  the direction of the proto-galactic 
angular momentum is on average  preferentially aligned with the 2nd 
principal axis of the shear tensor.  

For the ideal situation where the principal axis of the inertia tensor 
is totally independent of that of the shear tensor \cite{cat-the96},   
one can expect the maximal preferential alignment of the galaxy angular 
momentum vector along with the 2nd principal axis of the intrinsic shear 
tensor since 
$\langle I_{23}^2\rangle =\langle I_{13}^2\rangle =\langle I_{12}^2\rangle$.  
What has been found in LP00 numerical simulations is, however, 
far from  being idealistic.  
The principal axis of the inertia tensor has turned 
out to be quite strongly correlated with that of the shear tensor. 
However, a slight but detectable misalignment between the two 
tensors has been  detected by LP00. It means that in spite of 
the strong correlation between the inertia and shear tensors 
a net nonzero angular momentum at first order is indeed generated 
to a detectable degree with its axis preferentially aligned 
with the intermediate principal axis of the local shear. 
 
The essence of this idea is well represented by the following simple 
equation (see Appendix A): 
\beq
\langle \hat{L}_i \hat{L}_j |\hbT \rangle  =
\frac{1+a}{3}\delta_{ij}  - a\hat{T}_{ik} \hat{T}_{kj}. 
\label{eqn:lij}
\eeq 
Here $\hbT$ is a unit traceless local shear tensor 
($\hbT \equiv \tbT/|\tbT|$ 
where $\tT_{ij} \equiv T_{ij} - \delta_{ij}{\rm Tr}/3$), 
$\hbL$ is a unit galaxy spin vector, and $a$ is a 
{\it spin-shear correlation parameter} introduced by LP00 
to measure the strength of the correlation between the local 
shear and the galaxy spin axis.   
If $a= 0$, $\langle \hL_i \hL_j |\hbT \rangle = \delta_{ij}/{3}$, 
spins are randomly oriented without any correlation with the 
local shears.  While if the inertia and shear tensors are mutually 
uncorrelated, and there is no nonlinear effects, then 
the value of $a$ is calculated to be $3/5$ 
(it was mistakenly cited as unity in LP00, see Appendix A). 
The real value of $a$ should be determined empirically  
by numerical simulations, since in the linear theory 
one cannot estimate the strength of the correlation between 
the inertia and shear tensors from 1st principles, and one
expects non-linear effects to be important as well. LP00 suggested 
the formula for the estimation of $a$ from N-body simulations: 
\beq
a = 2 - 6\hlam_{i}^{2}\hL_{i}^{2}.     
\label{eqn:opa}
\eeq 
Here $\{\hlam_{i}\}_{i=1}^{3}$ are the three eigenvalues of 
the trace-free unit shear tensor, satisfying 
$\sum_{i}\hlam_{i}^2=1$ and $\sum_{i}\hlam_{i}=0$, while 
$\hbL$ is the unit angular momentum vector measured in the  
shear principal axis frame. 
Note that if equation (\ref{eqn:lij}) holds as a theoretical estimation 
for $\hL_{i}\hL_{j}$ , then equation (\ref{eqn:opa}) becomes optimal 
(see Appendix J).   
LP00 found $a \approx 0.24 $ in their N-body simulations. 
For the detailed description of the measurement of $a$ from N-body 
simulations used by LP00, see $\S 2$ in LP00. 
   
The numerical result of LP00 indicates that the present galaxy spin 
axes are indeed (weakly but detectably) correlated with the 
intrinsic shears even though the correlation is not very strong.  
It is worth noting that $a$ is a universal value, independent of scale.  
The spin-shear correlation parameter of $a$, by its definition, must be 
measured from the tidal shears smoothed on the same scale that 
$\hbL$ is defined on.  
 
Given the detectable preferential alignment of the galaxy spin 
along the 2nd principal axis of the intrinsic shear, it is possible to
reconstruct the shear field from the observable unit galaxy spins.
Let us say that we have $m$ galaxies with measured unit spins,
$\hbL(\bx_{\gamma})$ for $\gamma = 1,2,\cdots,m$.  Now, we would like
to find the maximum likelihood value of the traceless shear tensor,
$\tbT$ at each galaxy position.  Using Bayes' theorem,
$P(\tbT|\hbL)=P(\hbL|\tbT) P(\tbT)/P(\hbL)$.  An immediate
complication arises: $P(\hbL|\tbT)$ is a purely local process,
independent of the events at any other point, while
$P(\tbT)=P[\tbT(\bx_1),\tbT(\bx_2),\ldots]$ is a joint random process 
linking different points with one another.  As we noted in LP00, 
the linear shear expectation $\langle \tbT|\hbL\rangle = 0$ 
since $P(\hbL|\tbT)$ is an even function of $\tbT$.  Since we are using
only directions of the spins (which are more readily observed and
predicted), we cannot recover the magnitude of the shear field.  
In other words, the shear field can be reconstructed 
up to the ambiguity of a multiplicative normalization constant.  
Thus we can arbitrarily normalize the shear field.  
Here we use the normalization constraint of  
$\int \tT_{ij}(\bx)\tT_{ij}(\bx)d\bx = 1$.
The nontrivial quadratic maximum likelihood value of the shear field 
with this constraint is given as the solution to the following 
eigenvector equation 
(Appendix D):   
\beq
\int \txi_{ijlm}(\bx_\alpha,\bx_\beta)
\tT_{ij}(\bx_\alpha)d^3\bx_\alpha = \Lambda\tT_{lm}(\bx_\beta).   
\label{eqn:tlm}
\eeq
$\Lambda$ is the largest eigenvalue of the posterior correlation 
operator, \\ 
$\txi_{ijlm}(\bx_{\alpha},\bx_{\beta}) \equiv 
\langle\tT_{ij}(\bx_{\alpha})\tT_{lm}(\bx_{\beta})|\hbL\rangle$. 
In the asymptotic case of $a \ll 1$ as in LP00 simulation results,  
$\txi_{ijlm}(\bx_{\alpha},\bx_{\beta})$ is given (Appendix C) as 
\beq
\txi_{ijlm}(\bx_\alpha,\bx_\beta) = 
- a\int\tC_{ijnk}(\bx_\alpha-\bx_\gamma) 
\tC_{lmok}(\bx_\beta-\bx_\gamma) 
\hL_n(\bx_\gamma)\hL_o(\bx_\gamma)d^3\bx_\gamma,
\label{eqn:lin}
\eeq
where  $\tbC$ is a two-point covariance matrix of the traceless 
shear tensor defined as 
$\tbC = (\tC_{ijkl}) = \langle\tT_{ij}({\bf x})
\tT_{kl}(\bx+ \br)\rangle$ (see Appendix B). 

In Appendix D, we explain in detail using the Lagrange multiplier 
method that the eigenvector associated with the largest eigenvalue is 
indeed the maximum likelihood expectation value of the shear.   
It is worth mentioning that in practice the posterior 
correlation function is defined only accurately at each 
galaxy position, so the integral in equation (\ref{eqn:lin}) 
must be replaced by a sum over discrete galaxy positions.    
We have regarded small $a$ ($a \ll 1$) as the limit of small signal to 
noise.  We can also find a general expression for the 
shear reconstruction  in Fourier space (Appendix E):  
\beq
\int \frac{\txi_{ijlm}(\bk_\alpha,\bk_\beta)}{P(k_\alpha)
P(k_\beta)}\tT_{ij}(\bk_\alpha)d^3\bk_\alpha = 
\Lambda\tT_{lm}(\bk_\beta). 
\label{eqn:tpk}
\eeq
Here $P(k)$ is the density power spectrum.  
Note that equation (\ref{eqn:tpk}) is the optimal-filtered version 
of equation (\ref{eqn:tlm}), holding without the constraint  
of small $a$.  

Now, the expected shear field given the unit spin field can be found
as the eigenvector of $\tilde\xi_{ijlm}$ associated with the largest 
eigenvalue.  LP00 suggested an effective power iteration scheme to 
estimate the largest eigenvector of $\txi_{ijlm}$: 
One starts with an initial guess 
$\tilde{T}_{ij}^0$, and defines an iteration such that 
\ben 
\tT_{ij}^{n+1/2} &=& \int\tilde{\xi}_{ijlm} 
\tT^{n}_{lm}d^3\bk, \nonumber \\ 
\tT^{n+1}_{ij} &=& \frac{\tT^{n+1/2}_{ij}}
{\sqrt{\int(\tT_{ij}^{n+1/2})^2d^3\bk}} + \tT_{ij}^{n-1}.  
\een
Sufficiently large number of iterations converges the testing vector to 
the solution, i.e., the eigenvector associated with the largest 
eigenvalue with a small fractional estimation error proportional to 
$(\Lambda_1/\Lambda_0)^m$, where $m$ is the number of iterations, 
$\Lambda_0$ and $\Lambda_1$ are the largest and second largest
eigenvalues respectively.  Appendix F gives a general proof 
for the power iteration. 

In order to find the expected shear field by solving equation 
(\ref{eqn:tpk}) using  the above iteration method, one has to know 
the power spectrum of the mass density field beforehand. 
Here we describe how one  can actually determine the slope of the linear 
power spectrum in deriving the shear field by equation (\ref{eqn:tpk}):
From an observed set of $N$ galaxies and with an initial guess 
for the mass power spectrum, we construct a posterior shear 
correlation function (\ref{eqn:lin}), which is a $5N\times 5N$ matrix. 
From this posterior shear correlation function, one can construct 
a weighed posterior shear correlation function given in equation 
(\ref{eqn:tpk}), whose largest eigenvector is the shear field to be 
reconstructed by the iteration method described above.   
Here the largest eigenvalue $\Lambda$ is the likelihood.  
We can iterate this procedure itself to measure a self-consistent 
power spectrum by varying the power spectrum to maximize $\Lambda$.  
In other words, at the same time when one reconstructs the initial 
shear field, one can also measure the slope of the initial power spectrum 
by finding such power spectrum as maximize $\Lambda$ in equation 
(\ref{eqn:tpk}).  Note, however, that one can recover the slope but 
not the amplitude of the power spectrum since the likelihood $\Lambda$ 
in (\ref{eqn:tpk}) is independent of a multiplicative constant of 
the power spectrum.   

The final step is the reconstruction of the density field, $\delta(\bx)$
given the traceless shears, $\tbT(\bx_i)$ reconstructed at each galaxy 
position $\bx_i$. First we consider a orthonormal parametrization of 
the density and the five free components of the traceless shear tensor 
such that 
\ben 
y_0 &=& \frac{\delta}{\sqrt{3}},  \nonumber \\
y_1 &=& \frac{(-3-\sqrt{3})\tT_{11}+2\sqrt{3}\tT_{22}+
(3-\sqrt{3})\tT_{33}}{6},
\nonumber \\
y_2 &=&  \frac{(3-\sqrt{3})\tT_{11}+2\sqrt{3}\tT_{22}+
(-3-\sqrt{3})\tT_{33}}{6}, \nonumber \\
y_3 &=& \sqrt{2}\tT_{12}, \hspace{0.5cm} 
y_4 = \sqrt{2}\tT_{23}, \hspace{0.5cm}
y_5 = \sqrt{2}\tT_{31}.  
\een 
$\by$ in fact is a orthonormal vector-representation  
of the full shear, $\bT$ in terms of trace and traceless parts.  
Therefore the mutual correlation between the six components of $\by$ 
at the same position is always zero. 

Reconstructing $\delta(\bx)$ given $\hbT(\bx_i)$ amounts to 
finding $\langle y_{0}(\bx) | y_{1}(\bx_i), 
y_{2}(\bx_i), \cdots, y_{5}(\bx_i) \rangle$.   Since a linear 
combination of the Gaussian variables is also Gaussian, $\by$ is a 
Gaussian variable, and the covariance matrix of $\by$, say, 
$\bV$ ($V_{ij} \equiv \langle y_{i}y_{j}\rangle$) can be 
obtained by the linear transformation of the shear two-point 
correlations (eq. [\ref{eqn:scccor}]).  
We obtain the following expression for 
$\langle y_{0}(\bx) | y_{1}(\bx_i),y_{2}(\bx_i), 
\cdots, y_{5}(\bx_i) \rangle$ (see Appendix G):  
\beq 
\langle y_{0}(\bx) | y_{1}(\bx_i),y_{2}(\bx_i), 
\cdots, y_{5}(\bx_i) \rangle
= - \frac{U_{0\nu}y_{\nu}}{U_{00}},
\label{eqn:denc} 
\eeq
where $\bU \equiv \bV^{-1}$, and the Greek index, $\nu$ 
goes from $1$ to $5$.   
Equation (\ref{eqn:denc}) allows us to reconstruct the density 
field at an arbitrary spatial position $\bx$ once the traceless shear  
field is reconstructed at each galaxy position $\bx_i$. 
Note that the only mathematical complication that arises in the 
reconstruction algorithm is a matrix inversion. Therefore it is 
computationally tractable, involving only linear algebra. 
It is worth mentioning that although the density field is supposed 
to be reconstructed in Lagrangian space, the galaxy spins are 
measured in Eulerian redshift space.  We can regard this displacement 
between the Eulerian and Lagrangian spaces as noise, and convolve 
simply the two-point density correlation function, $\xi(r)$ with 
a Gaussian filter with a peculiar velocity dispersion 
$\sigma_v = 150$ km/s for spiral galaxies   
(see Davis, Miller, \& White 1997).    

\begin{figure}
\plotone{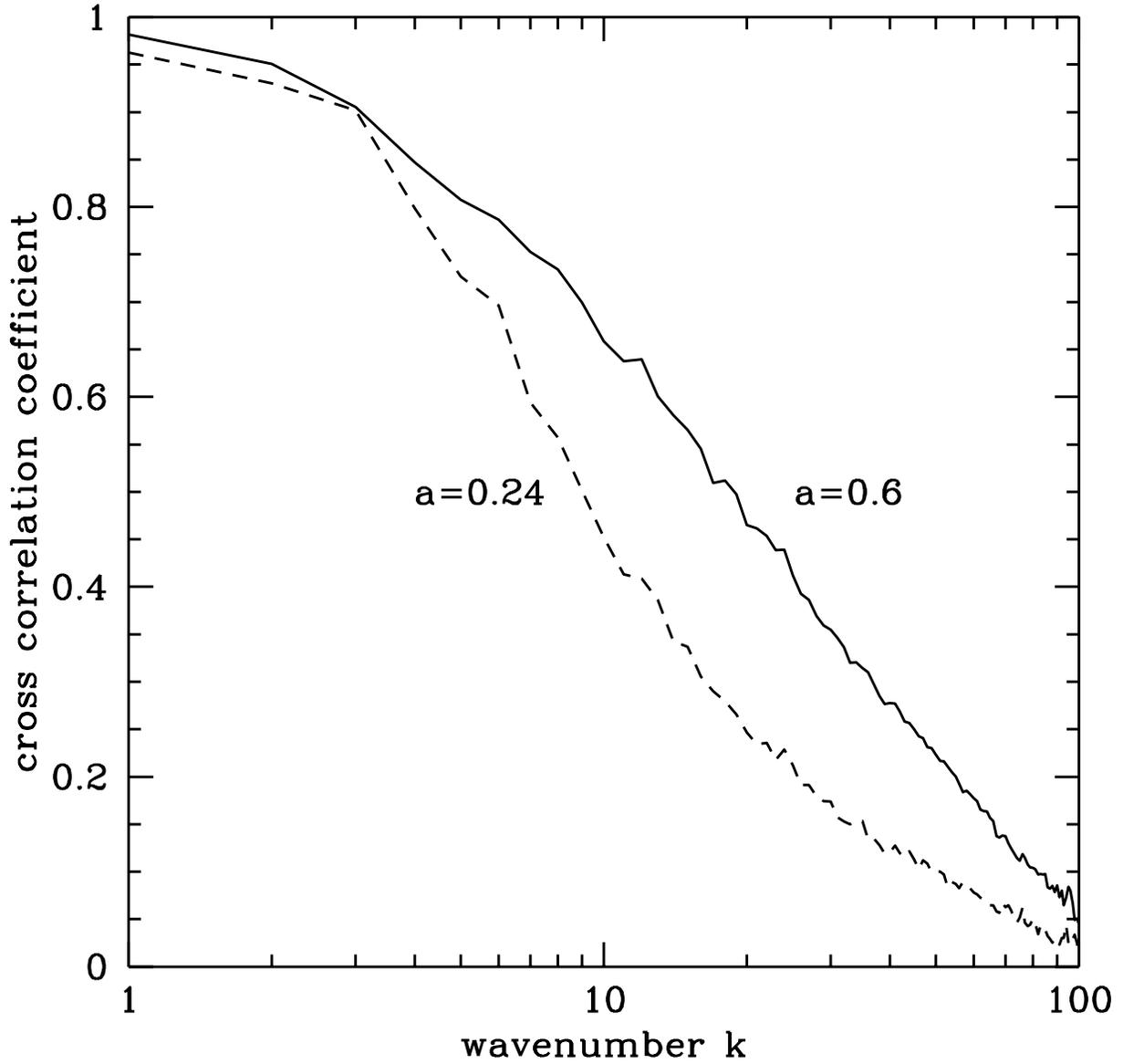}
\caption{Accuracy of the reconstructed density field.  The 
solid line is the ideal linear case, while the dotted line corresponds
to the strength of correlations measured in simulations and the Tully
data.  On large scales, reconstruction is always accurate. \label{fig1}}
\end{figure}

Fig.1 shows the accuracy of the reconstructed
density field.  It plots the cross correlation coefficient $r=\langle
\delta_r(k) \delta_m(k)\rangle/\sqrt{P_r(k) P_m(k)}$ between the
reconstructed density field $\delta_r$ and the true density field
$\delta_m$.  We have implemented the algorithm with the realistic 
value of the correlation parameter of $a = 0.24$ and simulated two  
million sample galaxy spins, and find a good ability of the algorithm 
to reconstruct the density field. 
As was shown in PLS00 and Crittenden \etal\ (2000a, 2000b), 
the intrinsic galaxy alignment is a much stronger effect
than weak lensing for shallow surveys such as SDSS. 
Given that it was hoped that weak lensing power spectra could
be measured, and thus the projected mass power spectrum, we would
clearly expect that this stronger effect of intrinsic alignments would
thus allow a better reconstruction of the density field in the source
plane.  The advantage now is that the full three dimensional shear
field can be reconstructed, not just a two dimensional projected 
field as would be the case for the weak lensing.  The above algorithm  
described in this section enables us to achieve this goal.

\section{SPIN-DENSITY CROSS CORRELATION} 

In order for the algorithm given in $\S 2$ to be applied to the real 
universe,  it is indispensable to have a nonzero shear-spin correlation 
parameter, $a$. Unfortunately, however, it is quite hard to measure 
the value of $a$ directly from real observational data since it requires 
us to know the initial shear field beforehand (see eq. [\ref{eqn:lij}]). 
An alternative simpler way to detect the intrinsic spin alignment  
is to investigate the spatial spin-spin correlation.  
The local shear effect is due to the surrounding matter distribution.   
But the matter in the universe is spatially correlated in the 
standard structure formation scenario.   
Consequently, if galaxy spins are indeed correlated with the 
linear shear tensor by equation (\ref{eqn:lij}) with a nonzero 
value of $a$, the spatial shear correlation must induce a 
spatial spin-spin correlation with themselves. 

Since galaxy formation is an unsolved problem, it is difficult to make 
an accurate quantitative evaluation of the expected level of the 
spin-spin correlation signal. We can make at most approximate analytic 
estimates for the order of magnitude of the spin 
correlation and its qualitative behavior.  
PLS00 have attempted to estimate the expected strength of the 
galaxy spin-spin correlation (see eq. [1] in PLS00) using the 1st-order 
perturbation theory and the numerical normalization amplitude of $a=0.24$.  
Appendix H lays out the detailed derivation of the spin-spin 
correlation function presented in PLS00.  PLS00 then measured the spin 
correlation signal directly from the observed spiral galaxies of the 
Tully catalog. The observed signal turned out to be significant at 
$97 \%$ confidence level, and the amplitude of the signal is of order of 
$1 \%$ at a separation of $1 h^{-1}$ Mpc, in agreement with the 
PLS00 theoretical estimates. 

The consistent results  of the observed spin-spin correlation signal 
with the theoretical predictions motivate us to consider 
a correlation of galaxy spin field with the density field.  
Galaxy spin alignment with the local gravitational shear field might 
result in the correlation of galaxy spins with the directional geometry 
of the nearby galaxy distribution.  Perhaps the simplest statistic to 
observe is the correlation between the spin axis $\hbL$ and the
spatial direction $\hbr$ to the nearest neighbor. 
Therefore, we first define a simple nontrivial spin-direction 
cross correlation function analogous to the spin-spin correlation 
function given by PLS00 such that : 
\beq
\omega(r)\equiv\langle |\hbL(\bx)\cdot\hbr(\bx)|^2\rangle -\omega_{0},  
\eeq
where $\omega_0$ is the value of $\omega(r)$ for the case of 
no correlation: $\omega_0 = 1/3$ for the three dimensional case  
while $\omega_0 = 1/2$ for the two dimensional case.  
Note that the two vectors, $\hbL$ and $\hbr$, are both defined 
at the same galaxy position, $\bx$.   

For a galaxy pair at $\bx$ and $\bx + \br$, let us consider the 
density and shear fields smoothed on two different top-hat scales, say 
$R$ and $R^{\pr}$, where $R$ is the top-hat galactic radius, while 
$R^{\pr}$ is the minimum top-hat radius that encloses the galaxy 
pair such that $R^{\pr} = R + r$.  In order to avoid confusion
about the smoothing scale, in this section we use an explicit notation 
of $\delta_{R}$ and  $\hbT^{R}$ respectively to represent the 
density and unit traceless shear fields smoothed on a scale of $R$, 
while $\delta_{R^{\pr}}$ and $\hbT^{R^{\pr}}$ for the density and 
unit traceless shear fields smoothed on a scale of $R^{\pr}$.     
A simple directional vector that one could form is the gradient of the
smoothed density field, $\nabla\delta_{R^{\pr}}(\bx)$.  Assuming that
galaxies form on peaks of the density field, one expects two
neighboring galaxies to sit at the ends of a ridge connecting the two
galaxies.  A peak is by definition a location where the gradient is
zero.  If one considers the gradient halfway between the two peaks,
one expects it to generically be near a saddle point, where again
the gradient is zero.  And in between, the gradient would be expected
to point in a direction perpendicular to the peak separation $\br$. 

Instead, one would expect that the direction of the galaxy 
separation vector correlates with the major principal axis of the 
local gravitational shear tensor smoothed on a scale of the galaxy 
separation.  If we neglect the other two principal axes (as is often 
the case in principal component analysis), we can relate 
$\hr_{i}\hr_{j}$ to $\hbT$ up to a considerable ambiguity.   
But, it results in a trivial spin-direction correlation:  
$\omega(r) \sim \langle\hT_{ik}\hT_{kj}\hT_{ij}\rangle = 0$ 
due to the even probability distribution of $\hbT$.  
We note, however, that the principal axis of a shear tensor is the 
same as that of its square, so we shall instead relate
$\hr_{i}\hr_{j}$ to $\hT^{R^{\pr}}_{ik}\hT^{R^{\pr}}_{kj}$ analogous to 
equation (3) such that  
\beq
\langle \hr_i \hr_j |\hbT^{R^{\pr}} \rangle  =
\frac{1 - b}{3}\delta_{ij}  + b\hT^{R^{\pr}}_{ik}\hT^{R^{\pr}}_{kj}, 
\label{eqn:rij}
\eeq  
where we introduce a new quantity, a {\it direction-shear correlation 
parameter} of $b$, to measure the strength of the correlation between 
the unit galaxy separation and the major axis of the local shear tensor.
A careful reader may have noticed the difference of the sign ahead 
of the correlation parameters  between equations (\ref{eqn:lij}) and 
(\ref{eqn:rij}).  This sign difference arises because the direction  
of each alignment with the unit shear tensor is different. $\hbr$ is 
aligned with the major principal axis of $(\hT_{ik}\hT_{kj})$ 
while $\hbL$ is aligned with the minor principal axis.   
Let $\lam_1,\lam_2,\lam_3$ be the three eigenvalues of the shear tensor, 
$\bT$ with the order of $\lam_1 > \lam_2 > \lam_3$. 
Then the three eigenvalues of the unit traceless shear tensor $\hbT$ 
is nothing but $\hlam_1, \hlam_2, \hlam_3$ with 
$\hlam_i = \lam_i - {\rm Tr}/3$, ${\rm Tr} = \sum_{i=1}^{3}\lam_i$.  
Obviously the order is the same: $\hlam_1 > \hlam_2 > \hlam_3$. 
Thus, the principal axes of $\bT$ and $\hbT$ coincide. 
But if we consider the square of the shear, $(\hT_{ik}\hT_{kj})$, 
the eigenvalues are given as $\hlam_{1}^{2},\hlam_{2}^{2},\hlam_{3}^2$ 
but with the order of $\hlam_{1}^{2} \sim \hlam_{3}^2 > \hlam_{2}^{2}$. 
Thus, the intermediate axis of the shear tensor becomes the minor 
axis of $(\hT_{ik}\hT_{kj})$. It explains why 
$\langle\hr_{i}\hr_{j}|\hbT\rangle$ is positively proportional to 
$\hT_{ik}\hT_{kj}$ (apart from the shear-independent constant) 
while $\langle\hL_{i}\hL_{j}|\hbT\rangle$ is negatively proportional to 
$\hT_{ik}\hT_{kj}$. Also note that in equation (\ref{eqn:lij}) $\hbT$ is 
smoothed on the top-hat galactic scale of $R$ while in equation 
(\ref{eqn:rij}) $\hbT$ is smoothed on the minimum enclosing 
top-hat radius of $R^{\pr}$ since the galaxy separation vector can be 
defined for a galaxy pair not for one galaxy.    
 
This direction-shear correlation parameter of $b$ can be also determined 
in N-body simulations in principle. We suggest the following 
formula for estimation of $b$ in simulations: 
\beq
b = \sqrt{2}\hlam_{i}\hr_{i}^{2}, 
\label{eqn:opb} 
\eeq 
where $\hbr$ is the unit separation vector in the shear principal axis 
frame.  In practice, each $\hr_i$ is obtained by measuring the separation 
vector of each closest galaxy pair and projecting the separation vector 
into the $i$-th principal axis of the local shears smoothed on the 
mean galaxy separation.  
Again equation (\ref{eqn:opb}) becomes optimal if equation (\ref{eqn:rij}) 
holds as a theoretical estimation formula for $\hr_{i}\hr_{j}$ 
(see Appendix J). 
We have found the average value of 
$b = 0.29 \pm 0.01$  from the same N-body simulation results 
that LP00 used for the measurement of $a$. 
It is worth mentioning, however, that the galaxy distribution is known 
to have a correlation function significantly different from that of the 
matter, measuring the value of $b$ requires a quantitative galaxy 
formation model beforehand.  Thus, with having no quantitative galaxy 
formation model,  equation (\ref{eqn:opb}) provides only a qualitative 
approximation for the magnitude of $b$.  
 
With the similar method that we have used for the spin-spin correlation, 
one can find an analytic estimates of $\omega (r)$ (Appendix I) such that:   
\beq
\omega (r) =  -A\frac{\langle\delta_{R^{\pr}}\delta_{R}\rangle^{2}}
{(\sigma_{R^{\pr}}\sigma_{R})^2},  
\label{eqn:dsw}      
\eeq
where the amplitude of $A$ depends on the correlation parameters, 
$a$ and $b$. 
It has the value of $ab/6$ and $5ab/24$ for the three and the 
two dimensional cases respectively.   
Here $\langle\delta_{R}\delta_{R^{\pr}}\rangle$ is the auto correlation 
of the density field smoothed  on two different scales of $R$ and $R^{\pr}$, 
and $\sigma_{R}$ and $\sigma_{R^{\pr}}$ are the corresponding rms density 
fluctuations:  
\ben 
\langle\delta_{R}(\bx)\delta_{R^{\pr}}(\bx)\rangle 
&=& \frac{1}{(2\pi)^3}\int W_{th}(kR) W_{th}(kR^{\pr}) P(k) 4\pi k^2 dk, 
\nonumber \\ 
\sigma^{2}_{R} =  
\langle\delta_{R}^{2}(\bx)\rangle &=&  
\frac{1}{(2\pi)^3}\int W^{2}_{th}(kR) P(k) 4\pi k^2 dk, \nonumber \\
\sigma^{2}_{R^{\pr}} = 
\langle\delta_{R^{\pr}}^{2}(\bx)\rangle &=&  
\frac{1}{(2\pi)^3}\int W^{2}_{th}(kR^{\pr}) P(k) 4\pi k^2 dk,   
\een
where  the top-hat window function is given as 
$W_{th}(kR) = 3[\sin(kR) - kR\cos(kR)]/(kR)^{3}$. 

To find a closed analytic form of $\omega (r)$, we can replace 
the top-hat filter with the Gaussian filter.  Using a Gaussian filter of 
$W_{G}(kR) = \exp(-k^{2}R^{2}/2)$, and a power-law power spectrum of
$P(k) = k^{-2}$, we find 
\beq
\omega (r) = -A\frac{2R^{\pr}R}{R^2 + {R^{\pr}}^2}.
\label{eqn:gsw}
\eeq
Equation (\ref{eqn:gsw}) says that for neighboring galaxies, 
$|\omega(r)|$ decreases as $r^{-1}$, less rapidly than the spin-spin 
correlation that decreases as $r^{-2}$ (see PLS00). Note that 
in Lagrangian space the galaxy separation of $r$ cannot decrease   
below $3R$ since the top-hat radius enclosing a galaxy pair must be 
at least three galactic scale radius of $R$. Thus we assign $\omega (r)$
a constant value of $\omega (3R)$ for $r \le 3R$. 

\section{SIGNAL FROM THE REAL UNIVERSE}

The unique and advantageous feature of the galaxy spin statistics 
presented in $\S 2$ and $\S 3$ is that it is a readily testable 
theory against real observational data since it deals not 
with the magnitude  but with the direction of a galaxy spin.  
The spin axis of a spiral galaxy can be easily determined by the  
information of the position angle (PA) and axial ratio ($\cal R$): 
A spiral galaxy is a thin disk with a circular face-on shape,  
and its spin vector is perpendicular to the plane of the disk.  
Therefore, the apparent axial ratio gives the magnitude of the 
radial component of a spin vector, while the position angle 
determines the relative magnitude of the tangential components 
of the spin vector lying in the plane of the sky. 

In order to apply observational tests to our theory, the most suitable 
dataset should be a large sample of spiral galaxies at low redshift 
with the information of position angle and axial ratio (${\cal R}$).  
The low redshift condition is required since at high redshift 
the weak lensing shear effect on the apparent orientation of the 
spin axis is dominant \citep{jai-sel97,wit-etal00,hea-etal00}. 
B. Tully (2000, private communication) generously has provided such a 
galaxy catalog: The Tully galaxy catalog is a compilation of 35674 nearby 
galaxy properties over the whole sky with median redshift of $6740$ km/s. 
Among the total 35674 Tully galaxy properties, 12122 galaxies are 
identified as spirals. 

In measuring the spin-direction correlation, we consider all the 
35674 galaxies in the Tully catalog to calculate the direction 
vectors, while we used only the spiral galaxies to measure the spin 
vectors. As mentioned in PLS00, we suspect that the shape-shape 
correlation of galaxies might cause a potential problem as a form of 
$\cal R$-related systematic errors.  The $\cal R$-related systematic 
errors are involved in the measurement of the axial ratio, ${\cal R}$,  
found in the Tully catalog, caused presumably by the deviation of the 
shape of spiral galaxies from a perfect ellipse, finite thickness of 
galaxies, and etc. For the detailed description of the Tully catalog 
and the data analysis, see $\S 3$ of PLS00. 

An easy way to avoid any false signal from the $\cal R$-related 
systematic errors is to measure the two dimensional spin-direction 
correlation. We project the three dimensional unit spin vector, 
$\hbL$ and unit separation vector, $\hbr$ onto the plane of the sky 
to obtain the two dimensional unit spin vector,  $\hbS = \bS/|\bS|$, 
$\bS = \hbL - (\hbL\cdot\hbx)\hbx$, and two dimensional 
separation vector, $\hbt = \bt/|\bt|$, $\bt=\br-(\hbr\cdot\hbx)\hbx$.
Now, the two dimensional spin-direction correlation is given by 
$\omega_{2D}(r) = \langle |\hbS\cdot\hbt |^{2}\rangle-1/2$. 
Note that the projection of the spin vector onto the plane of the sky 
amounts to setting ${\cal R}=0$, so $\omega_{2D}(r)$ is 
free of the $\cal R$-related systematic errors. 

For the three dimensional spin-direction correlation, we use an effective 
redistribution-method to deal with the $\cal R$-related systematic 
errors. We first bin the separation of every galaxy pair.  
At each bin we uniformly redistribute the given $\cal R$ of each galaxy 
spin in the range of $[0,1)$ and the radial component of the given 
$\hbr$ in range of $(-1,1)$ (the radial direction is along the 
line-of-sight at each galaxy position).  We expect that the uniform 
redistribution of $\cal R$ and $\hbr$ of galaxy pairs belonging to 
each bin eliminates effectively the systematic bias and false signal. 
Now we renormalize the spin and the separation vectors after the 
uniform redistribution of $\cal R$ and $\hbr$ at each bin, and 
calculate the three dimensional spin-direction correlation, 
$\omega_{3D}(r) = \langle |\hbL\cdot\hbr|^{2}\rangle - 1/3$.

\begin{figure}
\plotone{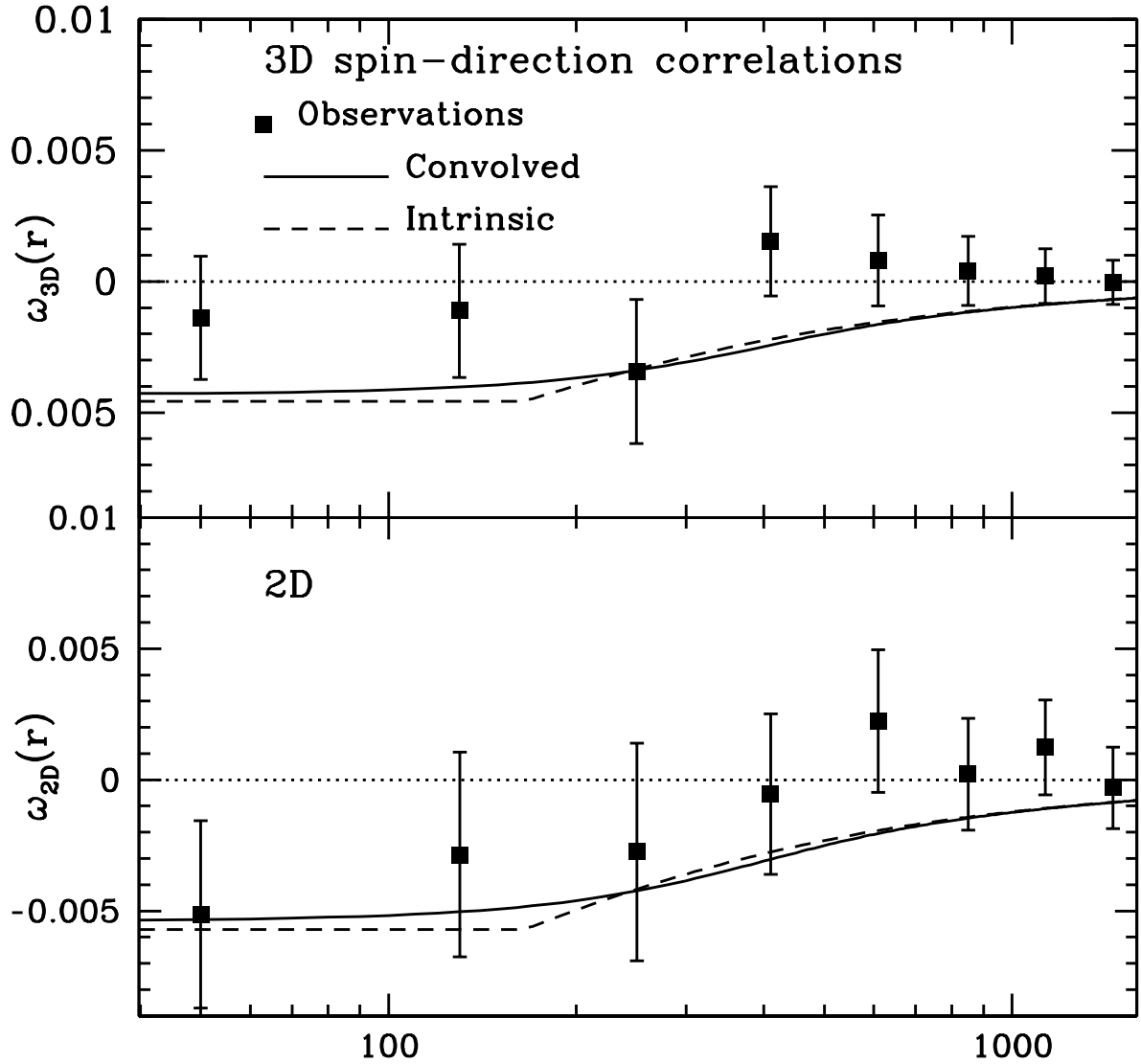}
\caption{The spin-direction correlation signal for 
the case of a power-law spectrum of $P(k) = k^{-2}$ with the 
correlation parameters having the value of  $a = 0.24$ and $b = 0.3$.
\label{fig2}}
\end{figure}

Fig. 2 plots the resulting observed signal (filled squares)   
vs. the galaxy separation, $r = cz$ (km/s) with error bars, and 
compares the observed signals with the theoretical estimates. 
Regarding the theoretical curves in Fig. 2, we convolve the 
Lagrangian correlation (dashed line) by a Gaussian filter with 
$\sigma_v = 150$ km/s to obtain the Eulerian correlation (solid line) 
the observed signal is measured in Eulerian redshift space. 
For a detailed description of the convolution procedure, see also 
$\S 3$ in PLS00.  The error bars are obtained from the experiment  
with the 500 sets of 12122 random two dimensional unit spins. 
We first generate the 12122 random spin vectors, and calculate the 
spin-direction correlation.   We repeat this process 1000 times 
with different sets of random spin vectors, and compute the standard 
deviation of the spin-direction correlations. The solid line 
is the theoretical predictions given by equation (\ref{eqn:dsw}) 
with the normalization amplitudes of $a = 0.24$ and $b = 0.3$ for the 
case of a power-law spectrum of $P(k) = k^{-2}$.  

Although the observed spin-direction correlation is fairly consistent 
with the theoretical estimates qualitatively, the signal is quite weak, 
and the error bars are still large.  We expect that larger surveys like 
SDSS will make a precision measurement of the spin-density correlation 
signal in the near future.

\section{SUMMARY AND CONCLUSIONS}

We have presented the technical formalism in which we discuss the
intrinsic galaxy spin correlation.  We have shown how the intrinsic
spin correlation is related to the initial potential and density
fields, and how the problem can be inverted to derive the power
spectrum and density field up to a multiplicative constant from the
observable orientation of galaxy spins, as originally claimed by LP00.
Since the intrinsic galaxy alignments are expected to dominate the
weak lensing signal for shallow surveys such as SDSS,  our algorithm
for the density reconstruction by the intrinsic galaxy spin alignment
should be more viable than the one by the weak lensing shear effect.

The formalism also allows us to address the issue of the spin-direction
correlation, which we have estimated theoretically, and measured in the
observational catalog.  Although the observed signal is reasonably
consistent with the theoretical estimates, the signal is quite weak,
and the error bars are still large.  We encourage future works on the
spin-direction correlation with larger surveys, which will make a
precision measurement of the spin-direction correlation signal.

\acknowledgements

We are very grateful to B. Tully for his catalog.
We also thank U. Seljak for useful discussions.   
This work has been supported by Academia Sinica and partially by  
NSERC grant 72013704 and computational resources of the 
National Center for Supercomputing Applications. 

\newpage

\appendix

\section{SHEAR-SPIN CORRELATION}

In order to find the expectation value of the unit galaxy angular momentum 
product given the unit shear tensor, $\langle \hL_i\hL_j |\hbT\rangle$,  
let us first consider $\langle L_i L_j |\bT\rangle$.  If the local  
shear tensor, $\bT$ and the inertia tensor, $\bI$  are mutually 
uncorrelated, the ensemble average of equation (\ref{eqn:1stl}) over all 
orientations of the inertia tensor gives 
\beq
\langle L_i L_j |\bT \rangle = \epsilon_{iab}\epsilon_{jcd}
T_{ak}T_{cl}\langle I_{kb}I_{ld}\rangle.
\label{eqn:ap1} 
\eeq
where the time-dependent proportionality constant in equation 
(\ref{eqn:1stl}) is set to be unity at present epoch. 

From the statistical isotropy of the inertia tensor,
we have the inertia tensor correlation 
$\langle I_{kb}I_{ld} \rangle =  \gamma (\delta_{kb}\delta_{ld}
 + \delta_{kl}\delta_{bd} + \delta_{kd}\delta_{bl})/3$ where 
$\gamma \equiv \langle (I_{11}^2 + I_{22}^{2} + I_{33}^{2})/3 \rangle $
is the proportionality constant. Here we stress that   the density 
reconstruction algorithm depends not on the magnitude but on the 
direction of $\bL$.  Thus, the overall proportionality constants that 
arise in the middle of our derivations can be always set to unity 
and we set $\gamma = 1$ hereafter. Of course, the 
renormalized angular momentum does not have the same magnitude 
as the original angular momentum. Furthermore, it does not have the 
dimension of the angular momentum anymore. But this renormalized vector 
does have the same direction as the original angular momentum, 
which is all that matters. Hereafter, we will use this kind of 
renormalization frequently by setting any proportionality constant to
be unity whenever it does not affect the direction of the angular 
momentum regardless of the dimensionality. 

Using $\langle I_{kb}I_{ld} \rangle =  (\delta_{kb}\delta_{ld}
 + \delta_{kl}\delta_{bd} + \delta_{kd}\delta_{bl})/3$, 
equation (\ref{eqn:ap1}) can be rewritten as  
\beq
\langle L_i L_j |\bT \rangle = \frac{\epsilon_{iab}\epsilon_{jcd}
\left( T_{ab}T_{cd} + T_{ad}T_{bd}\right) + 
\delta_{ij}|T|^2 - T_{ik}T_{kj}}{3}.
\label{eqn:ap2}
\eeq
On can verify that (\ref{eqn:ap2}) does not depend on the trace of 
the shear, so we rewrite it in terms of a traceless shear tensor  
$\tT_{ij}= T_{ij}-\delta_{ij}{\rm Tr}/3$ such that, 
\beq
\langle L_i L_j |\tbT \rangle  = 
\frac{2}{3}\delta_{ij}|\tT|^2 - \tT_{ik}\tT_{kj}.
\label{eqn:ap3}
\eeq 
Note that $\langle L_i L_i |\tbT \rangle = |\tbT|^2$ for this 
case of independent of $\bT$ and $\bI$.  Since we are again
only interested in the direction of the angular momentum vector, 
we may rescale equation (\ref{eqn:ap3}) to have the normalization 
constraint of $\langle L_i L_i |\hbT \rangle = 1$, dividing each side 
by $|\tbT|^2$ such that 
\ben
\langle L_i L_j |\hbT \rangle &=& 
\frac{2}{3}\delta_{ij} - \hT_{ik}\hT_{kj} \nonumber \\
&=& \frac{1}{3}\delta_{ij} + \left(\frac{1}{3}\delta_{ij} 
- \hT_{ik}\hT_{kj}\right). 
\label{eqn:ap4}
\een
The first term in the RHS of equation (\ref{eqn:ap4}) corresponds to  
the stochastic sources uncorrelated with the initial shear field. 
That is, the first term represents the ensemble average of 
$L_{i}L_{j}$ for the case that the direction of the angular momentum 
is completely random, having no correlation with the shear axis, 
due to the modification by the other stochastic sources such as 
nonlinear effect, mutual correlation between $\bI$ and $\bT$, and etc. 
While the second term in the RHS of equation (\ref{eqn:ap4}) corresponds 
to the deviation of the spin direction from the random 
average value due to it tendency to align preferentially with the 
intermediate axis of the shear tensor. 

This linear theory prediction for $\langle L_i L_i |\hbT \rangle$ 
holds provided that there is no nonlinear effects and that $\bT$ 
and $\bI$ are mutually independent. In practice, however, this 
condition is not guaranteed. Here we adopt the following simple 
assumption:  the non-linear and stochastic effects are uncorrelated
with the linear prediction, which adds noise to the unit spin vector.  
The nonlinear effect and the mutual dependence between $\bT$ and $\bI$
decreases the relative weight of the second shear-dependence term 
in equation (\ref{eqn:ap4}), which can be quantified by one parameter 
(say, $c$) such that 
\beq
\langle L_i L_j |\hbT \rangle  = 
\frac{1}{3}\delta_{ij} + c\left(\frac{1}{3}\delta_{ij} - 
\hT_{ik}\hT_{kj}\right).  
\label{eqn:ap5}
\eeq   
As the value of $c$ decreases, the first term of the RHS in 
equation (\ref{eqn:ap5}) dominates more, making the direction 
of the spin vector more random.  the linear theory predictions 
with  perfectly independent $\bI$ and $\bT$ corresponds to $c$ of unity, 
while the completely random direction of the angular momentum vector 
corresponds to $c = 0$.  
This is the generalized quadratic relation we propose to express 
the correlation between the direction of the present galaxy angular 
momentum vector and the initial gravitational shear tensor. 
Retaining the framework of the linear perturbation theory, we treat 
the existence of nonlinear effect and the correlation between 
the shear and the inertia tensors as stochastic sources 
that tend to randomize the direction of the spin vector,  decreasing 
the correlation of the angular momentum with the initial shear tensor. 

Now, to find the expression for $\langle \hL_i\hL_j |\hbT\rangle$, 
let us find the conditional probability density function, 
$P(\hbL|\tbT)$. The conditional probability density function, 
$P(\bL|\tbT)$ is usually given as a Gaussian distribution 
(see Catelan \& Theuns 1996) such that 
\beq
P(\bL|\tbT) =\frac{|\bQ|^{-1/2}}{\sqrt{(2\pi)^3}}
\exp\left(-\frac{\bL^{\rm T}\cdot\bQ^{-1}\cdot\bL}{2}\right), 
\label{eqn:ap6}
\eeq
where  the covariance matrix $\bQ$ is defined in equation
(\ref{eqn:ap5}).  The conditional probability density distribution  
$P(\hbL|\hbT)$ can be derived by integrating out $P(\bL|\tbT)$ over 
the magnitude of $L = |\bL|$ such that 
\beq
P(\hbL|\hbT) = \int P(\bL|\hbT)L^{2}dL = 
\frac{|\hbQ|^{-1/2}}{4\pi}\left(\hbL^{\rm T}\cdot
\hbQ^{-1}\cdot\hbL\right)^{-3/2}, 
\label{eqn:ap7}
\eeq
where $P(\hbL|\tbT)$ is in fact equal to $P(\hbL|\hbT)$. 
Here the unit covariance matrix, $\hQ_{ij}$ is given by equation 
(\ref{eqn:ap5}). 

In the limit of $c \ll 1$,  equation (\ref{eqn:ap7}) is 
simplified into 
\beq
P(\hbL|\hbT) = \frac{1}{4\pi}\left(1 + \frac{3c}{2}
[1 - 3\hT_{ik}\hT_{kj}\hL_{i}\hL_{j}]\right).
\label{eqn:ap9} 
\eeq
With equation (\ref{eqn:ap9}) and the help of little algebra, 
it is straightforward to calculate the expectation value of 
$\langle\hL_{i}\hL_{j}|\hbT\rangle$ in the limit of $c \ll 1$ 
such that 
\ben
\langle\hL_{i}\hL_{j}|\hbT\rangle &=& 
\int \hL_{i}\hL_{j} P(\hbL | \hbT) d\hbL, \nonumber \\
&=& \left(\frac{1}{3} + \frac{c}{5}\right)\delta_{ij}
 - \frac{3}{5}c\hT_{ik}\hT_{kj}. 
\label{eqn:ap10}
\een
Let us define a correlation parameter, $a \equiv 3c/5$.
Then, we finally get the desired expression :
\beq
\langle\hL_{i}\hL_{j}|\hbT\rangle
= \frac{1+a}{3}\delta_{ij}  - a\hat{T}_{ik} \hat{T}_{kj}.
\label{eqn:ap11}
\eeq
This equation says that for the ideal case of independent $\bI$ and $\bT$, 
the correlation parameter has the value of $a = 3/5$ 
(corresponding to $c = 1$) while for the random spins having no 
dependence on the shear tensor $a = 0$ ($c = 0$).

For practical purposes, it is also useful to have a similar expression 
to equation (\ref{eqn:ap11}) for the two dimensional unit spins. 
The two dimensional unit spins mean the galaxy spins projected onto 
the plane of sky, and normalized to have a unit magnitude. 
Let $(\hS_{1}, \hS_{2}) = (\cos\p, \sin\p )$ be the two dimensional 
unit spins, and $P(\hS_1, \hS_2|\hbT)d\hS = P(\p|\hbT)d\p$ be the 
conditional probability density distribution of the two dimensional 
unit spins.  Using the flat-sky approximation with $\hL_{3}$ in the 
line-of-sight direction, one can say 
$\hL_{1} = \sqrt{1 - \hL_{3}^2}\cos \p$, 
$\hL_{2} = \sqrt{1 - \hL_{3}^2}\sin \p$. 
Then, one can say $P(\hL | \hbT) = P(\p, \hL_{3} | \hbT)$. 
Now, $P(\p|\hbT)$ can be obtained by integrating out 
$P(\p, \hL_{3} | \hbT)$ over $\hL_{3}$ such that 
\ben
P(\p | \hbT) &=& \int^{1}_{-1}P(\p, \hL_{3} | \hbL) d\hL_{3} \nonumber \\ 
&=& \frac{1}{4\pi} \int^{1}_{-1}\left(1 + \frac{5a}{2}
[1 - 3\hT_{ik}\hT_{kj}\hL_{i}\hL_{j}]\right) d\hL_{3} \nonumber \\ 
&=& \frac{1}{2\pi}\bigg{\{}1 +  5a\left( \frac{1}{2} 
- \frac{f_3}{2} + f_{1}\cos^{2}\p + f_{2}\sin^2\p
- 2g_{12}\cos\p\sin\p\right)\bigg{\}},  
\label{eqn:2dd}
\een
where $f_i = \hT_{ik}\hT_{ki}$ for $i = 1,2,3$, and 
$g_{12} = \hT_{1k}\hT_{2k}$. 

Using equation (\ref{eqn:2dd}), it is straightforward to calculate 
$\langle \hS_i\hS_j|\hbT\rangle$ such that  
\ben
\langle \hS^2_{1}|\hbT\rangle &=& 
\int^{2\pi}_{0} \cos^2\p P(\p | \hbT) d\p, \nonumber \\
&=& \frac{1}{2} - \frac{5a}{8}(f_1 - f_2 ),
\label{eqn:2d1ij}
\een
\ben
\langle \hS^2_{2}|\hbT\rangle &=& 
\int^{2\pi}_{0} \sin^2\p P(\p | \hbT) d\p, \nonumber \\
&=& \frac{1}{2} - \frac{5a}{8}(f_2 - f_1 ). 
\label{eqn:2d2ij}
\een
\ben
\langle \hS_1\hS_2|\hbT\rangle &=& 
\int^{2\pi}_{0} \cos\p\sin\p P(\p | \hbT) d\p, \nonumber \\
&=& -\frac{5a}{4}g_{12}. 
\label{eqn:2d3ij}
\een

\section{SHEAR CORRELATIONS}

Let us calculate the spatial shear correlation, 
$C_{ijkl} = \langle T_{ij}(\bx) T_{kl}(\bx + \br) \rangle$.  
Using $T_{ij} = \partial_i\partial_j\phi$ and 
$\phi = \nabla^{-2}\delta$, one can write 
\ben 
C_{ijkl}(\br) &=& \langle T_{ij}(\bx) T_{kl}(\bx + \br) \rangle \nonumber \\  
&=& \langle \partial_{x_i}\partial_{x_j}\nabla_{\bx}^{-2}\delta(\bx)
\partial_{r_k}\partial_{r_l}\nabla_{\br}^{-2}\delta(\bx+\br) \rangle.
\label{eqn:sc1}
\een

Replacing the ensemble average with the spatial average by   
the ergodic theorem, and applying the integration by parts, 
one can show that equation (\ref{eqn:sc1}) 
can be rewritten as
\ben
C_{ijkl}(\br) = \partial_i\partial_j\partial_k\partial_l 
\nabla_{\br}^{-4}\xi(r) 
\label{eqn:cijkl1}
\een 

Using the identity relation, 
$\nabla_{\br}^{-2} = \int^{r}dr^{\pr}(1/r^{\pr2})
\int^{r^{\pr}}dr^{\pr\pr}r^{\pr\pr 2}$, 
and with the help of little algebra, equation (\ref{eqn:cijkl1}) 
can be arranged such that
\ben  
C_{ijkl}({\bf r}) &=&
(\delta_{ij}\delta_{kl}+\delta_{ik}\delta_{jl}+\delta_{il}\delta_{jk})
\bigg{\{}\frac{J_3}{6} - \frac{J_5}{10}\bigg{\}} + 
(\hat{r}_i\hat{r}_j\hat{r}_k\hat{r}_l)
\bigg{\{}\xi(r) + \frac{5J_3}{2} - \frac{7J_5}{2}\bigg{\}}
 \nonumber \\ 
&& +  (\delta_{ij}\hat{r}_k\hat{r}_l + \delta_{ik}\hat{r}_j\hat{r}_l
+ \delta_{il}\hat{r}_k\hat{r}_j + \delta_{jk}\hat{r}_i\hat{r}_l
+ \delta_{jl}\hat{r}_i\hat{r}_k + \delta_{kl}\hat{r}_i\hat{r}_j )
\bigg{\{} \frac{J_5}{2}-\frac{J_3}{2} \bigg{\}},    
\label{eqn:scccor}
\een
where $\hat{\bf r}={\bf r}/r$, 
$J_n \equiv n r^{-n}\int_0^r \xi(r^{\pr}) r^{\pr n-1} dr^{\pr n-1}$.
The two-point covariance matrix of the traceless shears 
$\tbC$  can  be also obtained by $\tC_{ijkl} = C_{ijkl} - 
\delta_{kl}C_{ijnn}/3 - \delta_{ij}C_{mmkl}/3 + 
\delta_{ij}\delta_{kl}C_{mmnn}/9$.

\section{POSTERIOR CORRELATION FUNCTION}

Let us consider the traceless posterior correlation function, \\ 
$\txi_{ijlk}(\bx_{\alpha},\bx_{\beta}) \equiv 
\langle \tT_{ij}(\bx_{\alpha})\tT_{lm}(\bx_{\beta})|
\hbL(\bx_{\gamma}) \rangle = 
\langle \taT_{ij}\tbeT_{lm}|\hbrL \rangle$ 
in the asymptotic limit of $a \ll 1$.  Here $\bx_{\alpha}$ and 
$\bx_{\beta}$ are two fixed galaxy positions where we would like 
to reconstruct the shear field, while $\bx_{\gamma}$ represents  
any arbitrary position of the given $N$ galaxies such that the index 
$\gamma = 1, 2,\cdots, N$ is dummy. 
Thus, it is in fact the expectation value of the quadratic shears 
given the whole galaxy spin field:     
\ben
\langle \taT_{ij}\tbeT_{lm}|\hbrL \rangle 
&=& \int d\tbaT\int d\tbbT\  \taT_{ij}\tbeT_{lm} 
P(\tbaT,\tbbT|\hbrL), \nonumber \\ &=& 
\int {\cal D}\tbrT\int d\tbaT\int d\tbbT\ \taT_{ij}\tbeT_{lm} 
P(\tbaT,\tbbT,\tbrT|\hbrL), \nonumber \\
&=& \int {\cal D}\tbrT\int d\tbaT\int d\tbbT\ \taT_{ij}\tbeT_{lm}
P(\tbaT,\tbbT,\tbrT)\frac{P(\hbrL|\tbaT,\tbbT,\tbrT)}{P(\hbrL)}, 
\nonumber \\ 
&=& \int {\cal D}\tbrT\int d\tbaT\int d\tbbT\ \taT_{ij}\tbeT_{lm}
P(\tbaT,\tbbT,\tbrT)P(\hbrL|\tbrT), 
\label{eqn:pcf1}
\een
where ${\cal D}\tbrT \equiv \prod_{\gamma=1}^{N}d\tbrT$, 
$P(\tbaT,\tbbT,\tbrT|\hbrL) = P(\tbaT,\tbbT,\tbrT)
P(\hbrL|\tbaT,\tbbT,\tbrT)/P(\hbrL)$ by Bayes's theorem. 

Here we use the approximation of 
$P(\hbrL) = P(\hbL^{1},\cdots,\hbL^{N}) 
\approx \prod_{\gamma=1}^{N}P(\hbrL) = 1/(4\pi)^{N}$. 
This approximation can be justified as follows: 
Let $P(\hbL^{1},\hbL^{2})$ be the joint probability distribution 
of the galaxy angular momentum. One can show that this joint 
probability distribution can be written as 
$P(\hbL^{1},\hbL^{2}) = \prod_{\gamma=1}^{2}P(\hbL^{\gamma}) + 
{\cal O}(a^{2})$ (see Appendix H). 
Thus, in the limit of $a \ll 1$, this approximation 
holds at first order of $a$.  
In this asymptotic limit, we also have $P(\hbrL|\tbaT,\tbbT,\tbrT) 
= P(\hbrL|\tbrT)$ using that the correlation of $\hbL$ with $\tbT$ at 
different points is also ${\cal O}(a^2)$ which we neglect.   
The constant, $\prod_{\gamma=1}^{N}P(\hbrL) = 1/(4\pi)^{N}$ is 
rescaled to unity in equation (\ref{eqn:pcf1}) since the exact 
value of the overall constant is irrelevant to the shear reconstruction 
(any positive proportionality constant can be rescaled to unity).  

Furthermore, equation (\ref{eqn:ap9}) says that in this asymptotic 
limit of $a \ll 1$, $\hbL$-dependent part of $P(\hbrL|\tbrT)$ 
is given as 
\beq
P(\hbrL|\tbrT) = -a\trT_{nk}\trT_{ko}\hrL_{n}\hrL_{o},  
\label{eqn:pcf2} 
\eeq
apart from a proportionality constant ($c = 5a/3$).  
Here the $\hbL$-independent part of $P(\hbrL|\tbrT)$ is 
ignored since it does not affect the shear-reconstruction, either. 
Inserting equation (\ref{eqn:pcf2}) into equation (\ref{eqn:pcf1}) gives 
\ben
\txi_{ijlm}(\bx_{\alpha},\bx_{\beta}) 
&=& -a\int {\cal D}\tbrT 
\int d\tbaT\int d\tbbT\ \taT_{ij}\tbeT_{lm}\trT_{nk}\trT_{ko}
P(\tbaT,\tbbT,\tbrT)\hrL_{n}\hrL_{o},\nonumber \\  
&=& -a\int d\bx_{\gamma}\langle\taT_{ij}\tbeT_{lm}\trT_{nk}\trT_{ko}
\rangle\hrL_{n}\hrL_{o}, \nonumber \\
&=& -a\int d\bx_{\gamma}\ \tC_{ijnk}(\bx_{\alpha}-\bx_{\gamma})
\tC_{lmok}(\bx_{\beta}-\bx_{\gamma})
\hL_{n}(\bx_{\gamma})\hL_{o}(\bx_{\gamma}).
\label{eqn:wich}
\een
Here $\langle\prod_{\gamma}\taT_{ij}\tbeT_{lm}\trT_{nk}\trT_{ko}\rangle 
=\sum_{\gamma}\langle\taT_{ij}\trT_{nk}\rangle\langle\tbeT_{lm}\trT_{ko}
\rangle =\sum_{\gamma}\tC_{ijnk}(\bx_{\alpha}-\bx_{\gamma})
\tC_{lmok}(\bx_{\beta}-\bx_{\gamma})$ by the Wick theorem.  
We ignore the other term  
$\langle\taT_{ij}\tbeT_{lm}\rangle\langle\trT_{nk}\trT_{ko}\rangle$ 
since this term does not depend on the distance of 
$\bx_{\alpha}-\bx_{\gamma}$ (or $\bx_{\beta}-\bx_{\gamma}$), having  
no contribution to the shear reconstruction through the galaxy spins. 
In the continuum limit, the sum is replaced by the integration 
over $\bx_{\gamma}$.

\section{MAXIMUM LIKELIHOOD EXPECTATION VALUE}

In this appendix, we provide a general argument 
that the maximum likelihood expectation value of a Gaussian 
random field can be given as the eigenvector associated with the 
maximum eigenvalue of the corresponding covariance matrix.

Let $\bfv$ be a Gaussian random field. Then, the probability 
distribution $P(\bfv)$ is Gaussian proportional to 
$\exp(-\bfv^{T}\cdot\bA^{-1}\cdot\bfv/2)$ where $\bA$ is the covariance 
matrix of $\bfv$.  Provided that $\bA$ is positive definite, 
the maximum likelihood value  of $\bfv$ must be the one that 
maximizes $\exp(-\bfv^{T}\bA^{-1}\bfv/2)$ or equivalently the 
one that minimizes $(\bfv^{T}\cdot\bA^{-1}\cdot\bfv)/2$. 
There is an obvious trivial solution, $\bfv = 0$ for all points, 
which is of course not the solution to be sought for. 

A nontrivial solution can be found by imposing a constraint.
Let us choose a quadratic constraint of $\bfv^{T}\cdot\bfv = 1$. 
Then, using the {\it Lagrange multiplier} method, we can 
say that the solution, i.e., the maximum likelihood value of $\bfv$ 
under this constraint should satisfy the following equation:  
\beq
\frac{\delta}{\delta\bfv}\left(\frac{\bfv^{T}\cdot\bA^{-1}\cdot\bfv}{2} 
- \frac{\lambda}{2}[\bfv^{T}\cdot\bfv - 1]\right) = 0, 
\label{eqn:lmm} 
\eeq
where $\lambda$ is a Lagrange multiplier. 

Solving the above equation gives
\beq
\bA^{-1}\cdot\bfv = \lambda\bfv,\hspace{0.5cm}\bfv^{T}\cdot\bfv = 1.
\label{eqn:lam}
\eeq 
Equation (\ref{eqn:lam}) says that the solution to equation 
(\ref{eqn:lmm}) is the eigenvector of $\bA^{-1}$ with the associated 
eigenvalue of $\lambda$. Thus, the eigenvector of $\bA^{-1}$ 
associated with the smallest eigenvalue minimizes   
$(\bfv^{T}\cdot\bA^{-1}\cdot\bfv)/2$, since 
$\bfv^{T}\cdot\bA^{-1}\cdot\bfv = \bfv^{T}\cdot\lambda_{min}\bfv 
= \lambda_{min}$. 
But, the eigenvector of $\bA^{-1}$ associated with the eigenvalue of 
$\lambda$  is also the eigenvector of $\bA$ itself associated with 
the eigenvalue, $1/\lambda$.  Therefore,  the eigenvector of 
$\bA^{-1}$ associated with the smallest eigenvalue, $\lambda_{min}$ 
is in fact the eigenvector of $\bA$ associated with the {\it largest}
eigenvalue, $1/\lambda_{min} \equiv \Lambda_{max}$.

Hence, one can say that the maximum likelihood expectation value 
of $\bfv$ is in fact the eigenvector of the positive definite 
covariance matrix of $\bA$ associated with the largest eigenvalue. 

\section{INVERSION THEOREM}

This Appendix is devoted fully to prove equation (\ref{eqn:tpk}),  
a nontrivial mathematical theorem ({\it inversion theorem}), 
which is at the core of our density reconstruction procedure.   

The inversion theorem says the following: If a unit galaxy spin is 
related to a unit traceless intrinsic shear tensor by equation 
(\ref{eqn:lij}) with a nonzero value of $a$, then it is possible  
to invert the measurable unit galaxy spin field into the initial 
intrinsic shear field by equation (\ref{eqn:tpk}). In other words, 
given the unit spin field, the expected intrinsic shear field is the 
solution to equation (\ref{eqn:tpk}) as the eigenvector 
associated with the largest eigenvalue of the posterior 
correlation function defined in equation (\ref{eqn:lin}).
  
In order to prove this inversion theorem, we first prove the 
following three lemmas.

{\bf Lemma 1}:   
\beq 
A_{lm} \equiv \sum_{i,j}\hT_{ij}\hT_{jl}\hT_{im} 
\Longrightarrow \tA_{lm}= \frac{\hT_{lm}}{2}.
\label{eqn:lem1}  
\eeq
In proving Lemma 1, we do not use the Einstein summation rule, so that 
the repeated indices do not mean the summation in the following proof 
(but the Einstein summation rule will be recovered after this Lemma 1).  
Let us first consider the off-diagonal elements, $A_{lm}$ with $l\neq m$.   
\ben
\tA_{lm} 
&=& A_{lm} = \sum_{i,j}\hT_{ij}\hT_{jl}\hT_{im} \nonumber \\
&=&\sum_{j}\hT_{lj}\hT_{jl}\hT_{lm} + 
\sum_{j}\hT_{mj}\hT_{jl}\hT_{mm} + 
\sum_{i\neq l, i\neq m, j}\hT_{ij}\hT_{jl}\hT_{im}
\label{eqn:lep11}  
\een
Note that the above equation is correct only in the three dimensional  
case where there is only one choice among $1$,$2$, $3$ for 
the dummy index $i$, if $i \neq l$ and $i \neq m$. So, in the final 
term of equation (\ref{eqn:lep11}), the index $i$ is not dummy. 
Since ${\rm Tr}({\hbT}) = 0$, we have
\ben
\sum_{j}\hT_{ij}\hT_{jl} &=& -\hT_{il}\hT_{mm} + \hT_{im}\hT_{ml}, 
\nonumber \\
\sum_{j}\hT_{mj}\hT_{jl} &=& -\hT_{ii}\hT_{ml} + \hT_{im}\hT_{il}.
\label{eqn:lep12}  
\een 
Using the above equations, one can say
\ben
&&\sum_{j}\hT_{ij}\hT_{jl}\hT_{im} + \sum_{j}\hT_{mj}\hT_{jl}\hT_{mm}
\nonumber \\
&=& \left(-\hT_{il}\hT_{mm} + \hT_{im}\hT_{ml}\right)\hT_{im} + 
\left(-\hT_{ii}\hT_{ml} + \hT_{im}\hT_{il}\right)\hT_{mm} 
\nonumber \\
&=& \left(\hT_{im}^2 - \hT_{ii}\hT_{mm}\right)\hT_{ml}
\label{eqn:lep13}  
\een
Thus, we have
\ben
\tA_{lm} &=& \sum_{j}\hT_{lj}\hT_{jl}\hT_{lm} + 
\left(\hT_{im}^2 - \hT_{ii}\hT_{mm}\right)\hT_{ml} \nonumber \\
&=& \left(\hT_{il}^2 + \hT_{ll}^2 + \hT_{lm}^2 + \hT_{im}^2 
- \hT_{ii}\hT_{mm}\right)\hT_{lm} \nonumber \\
&=& \left(\hT_{il}^2 + \hT_{lm}^2 + \hT_{im}^2 + \hT_{ii}^2 + 
\hT_{mm}^2 + \hT_{ii}\hT_{mm}\right)\hT_{lm} \nonumber \\
&=& \frac{1}{2}\hT_{lm},
\label{eqn:lep14}
\een
since 
$\hT_{ll}^2 = \hT_{ii}^2 + \hT_{mm}^2 + 2\hT_{ii}\hT_{mm}$ and 
$|\hbT|^2 = 1$
With the exactly same manner, one can also prove for the diagonal 
element, $\tA_{ll} = \hT_{ll}/2$. 

{\bf Lemma 2:}
\beq
\tC_{ijlm}(r) \equiv \langle\tT_{ij}({\bx})\tT_{lm}({\bx} +{\br})
\rangle \Longrightarrow 
\tC_{ijkl}(k) = \left(\hk_{i}\hk_{j} - \frac{\delta_{ij}}{3}\right)
\left(\hk_{l}\hk_{m} - \frac{\delta_{lm}}{3}\right) P(k), 
\label{eqn:lem2}   
\eeq
where $P(k) = |\delta_k|^2$ is the density power spectrum.

By the convolution theorem, we have 
\ben
\tC_{ijlm}(k) = \tT_{ij}({\bk})\tT_{lm}^{*}(\bk) 
&=& \left(k_{i}k_{j} - \frac{\delta_{ij}}{3}k^2\right)
\left(k_{l}k_{m} - \frac{\delta_{lm}}{3}k^2\right)|\Phi|^{2} \nonumber \\
&=& \left(\hk_{i}\hk_{j} - \frac{\delta_{ij}}{3}\right)
\left(\hk_{l}\hk_{m} - \frac{\delta_{lm}}{3}\right)P(k) 
\label{eqn:lep21} 
\een
since $T_{ij}(k) = k_{i}k_{j}\Phi(k)$, ${\rm Tr}(\bT) = \delta$, 
and $\delta(k) = k^2\Phi(k)$. 

{\bf Lemma 3:}
\beq
\frac{\tC_{ijml}(k)}{P(k)}\tT_{lm}(k) = \frac{2}{3}\tT_{ij}(k).
\label{eqn:lem3}  
\eeq 

\ben
\frac{\tC_{ijml}(k)}{P(k)}\tT_{lm}(k) 
&=& \left(\hk_{i}\hk_{j} - \frac{\delta_{ij}}{3}\right)
\left(\hk_{l}\hk_{m} - \frac{\delta_{lm}}{3}\right)T_{lm}(k) 
\nonumber \\
&=&\left(\hk_{i}\hk_{j} - \frac{\delta_{ij}}{3}\right)
\frac{2}{3}k^{2}|\Phi(k)|^2 \nonumber \\
&=& \frac{2}{3}\tT_{ij}(k).
\label{eqn:lep31} 
\een

Now, we are ready to prove the inversion theorem with the help of 
the above three lemmas.  From here on,  we will regard all the 
proportionality constants as unity. We will use equation (\ref{eqn:lij}) 
as a theoretical estimation formula for $\hL_{i}\hL_{j}$,  
discarding the shear independent $\delta_{ij}$-term in equation 
(\ref{eqn:lij}) since it does not affect the shear inversion:

{\bf Inversion Theorem:}
\beq
\hL_{i}({\bx})\hL_{j}({\bx}) = -\hT_{il}({\bx})\hT_{lj}({\bx}) 
\Longrightarrow 
\int \frac{\txi_{abcd}({\bk}_{\alpha},{\bk}_{\beta})}{P(k_{\alpha})
P(k_{\beta})}\tT_{ab}({\bk}_{\alpha})d{\bk}_{\alpha} = 
\tT_{cd}({\bk}_{\beta}), 
\label{eqn:inv}
\eeq
where
\beq
\txi_{abcd}({\bx}_{\alpha},{\bx}_{\beta}) 
= -\int\tC_{abil}({\bx}_{\alpha}-{\bx}_{\gamma})\tC_{cdjl}
({\bx}_{\beta}-{\bx}_{\gamma})
\hL_{i}({\bx}_{\gamma})\hL_{j}({\bx}_{\gamma})d{\bx}_{\gamma}.
\label{eqn:inp1}
\eeq

By the convolution theorem, 
\beq
\txi_{abcd}({\bk}_{\alpha},{\bk}_{\beta}) = 
-\tC_{abil}({\bk}_{\alpha})\tC_{cdjl}({\bk}_{\beta})\int\hL_{i}
({\bk}_{\alpha}+{\bk}_{\beta}-{\bk}^{\pr})
\hL_{j}({\bk}^{\pr})d{\bk}^{\pr}. 
\label{eqn:inp2}
\eeq
Now, by Lemma 3, we have
\ben
&&\int \frac{\txi_{abcd}({\bk}_{\alpha},{\bk}_{\beta})}{P(k_{\alpha})
P(k_{\beta})}\tT_{ab}({\bk}_{\alpha})d{\bk}_{\alpha}, \nonumber \\
&=& -\int\frac{\tC_{cdjl}({\bk}_{\beta})}{P(k_{\beta})}
\frac{\tC_{cdil}({\bk}_{\alpha})}{P(k_{\alpha})}\tT_{ab}({\bk}_{\alpha})
\int\hL_{i}({\bk}_{\alpha}+{\bk}_{\beta}-{\bk}^{\pr})
\hL_{j}(k^{\pr})
d{\bk}^{\pr}d{\bk}_{\alpha}, \nonumber \\
&=& -\int\frac{\tC_{cdjl}({\bk}_{\beta})}{P(k_{\beta})}
\tT_{il}({\bk}_{\alpha})
\int\hL_{i}({\bk}_{\alpha}+{\bk}_{\beta}-
{\bk}^{\pr})\hL_{j}({\bk}^{\pr})
d{\bk}^{\pr}d{\bk}_{\alpha}. 
\label{eqn:inp3}
\een 
Let $H_{ij}({\bx})=\hL_{i}({\bx})\hL_{j}({\bx})=
-\hT_{in}({\bx})\hT_{nj}({\bx})$.
Then, by the convolution theorem, $H_{ij}({\bk}_{\alpha}+{\bk}_{\beta}) 
= \int\hL_{i}({\bk}_{\alpha}+{\bk}_{\beta}-
{\bk}^{\pr})\hL_{j}({\bk}^{\pr})d{\bk}^{\pr}$. 
So equation (\ref{eqn:inp3}) is written as  
\beq
-\int\frac{\tC_{cdjl}({\bk}_{\beta})}{P(k_{\beta})}\tT_{il}({\bk}_{\alpha})
H_{ij}({\bk}_{\alpha}+{\bk}_{\beta})d{\bk}_{\alpha} = 
-\frac{\tC_{cdjl}({\bk}_{\beta})}{P(k_{\beta})}\int
\tT_{il}({\bk}_{\alpha})H_{ij}({\bk}_{\alpha}+{\bk}_{\beta})d{\bk}_{\alpha}.
\label{eqn:inp4} 
\eeq 
Let us define 
$A_{jl}({\bx}) \equiv  H_{ij}({\bx})\tT_{il}({\bx}) = 
-|\tbT|\hT_{in}({\bx})\hT_{nj}({\bx})\hT_{il}({\bx})$. 
Then in Fourier space one can say  
$A_{jl}({\bk}_{\beta}) =  
\int\tT_{il}({\bk}_{\alpha})H_{ij}({\bk}_{\alpha}+{\bk}_{\beta})
d{\bk}_{\alpha}$ by the convolution theorem.   
Let us decompose $\bA$ into the trace-free part and trace part 
such that $A_{jl}(\bx) = \tA_{jl}(\bx) + \delta_{jl}{\rm Tr}(\bA)/3$.   
But we already know from Lemma 1, the trace-free part of 
$\hT_{in}({\bx})\hT_{nj}({\bx})\hT_{il}({\bx})$ is given as 
$\hT_{jl}({\bx})$ (apart from the proportionality constant).  
Therefore, $\tA_{jl}(\bx) = -|\tbT|\hT_{jl}({\bx}) = -\tT_{jl}(\bx)$. 
In Fourier space 
$\tA_{jl}({\bk}_{\beta})  = - \tT_{jl}({\bk}_{\beta})$. 
Thus, equation (\ref{eqn:inp4}) becomes
\ben
-\frac{\tC_{cdjl}({\bk}_{\beta})}{P(k_{\beta})}\int
\tT_{il}({\bk}_{\alpha})H_{ij}({\bk}_{\alpha}+{\bk}_{\beta})d{\bk}_{\alpha}
&=& -\frac{\tC_{cdjl}({\bk}_{\beta})}{P(k_{\beta})}A_{jl}({\bk}_{\beta}) 
\nonumber \\ 
&=& -\frac{\tC_{cdjl}({\bk}_{\beta})}{P(k_{\beta})}
\left( \tA_{jl}({\bk}_{\beta}) + \frac{\delta_{jl}}{3}{\rm Tr}(\bA)
\right) \nonumber \\
&=& \frac{\tC_{cdjl}({\bk}_{\beta})}{P(k_{\beta})}
\tT_{jl}({\bk}_{\beta}),
\label{eqn:ctp}  
\een
since $\tC_{cdjl}\delta_{jl} = 0$ by equation (\ref{eqn:scccor}). 

But, by Lemma 3, equation (\ref{eqn:ctp}) is equal to 
$\tT_{cd}({\bk}_{\beta})$, which finally proves the inversion theorem. 

\section{POWER ITERATION}

We will provide a general proof for the power iteration scheme in 
this appendix.   

Let us assume that we have a real symmetric positive definite $n\times n$ 
matrix, $\bA$, and we seek for the eigenvector associated with the largest 
eigenvalue of $\bA$. Let us say, $\bfv_1, \bfv_2, \cdots, \bfv_n$ are the  
the $n$ eigenvectors of $\bA$ with the associated eigenvalues of 
$a_1, a_2, \cdots, a_n$ respectively 
(here we assume $a_1 \ge a_2 \cdots \ge a_n \ge 0$). 
If $n$ is not too large, then we can always find the eigenvectors along   
with the associated eigenvalues by solving the eigenvector equation, 
$\bA\bfv_i = a_i\bfv_i$ numerically. 
However, in the case $n$ is very large, finding the maximum eigenvector 
by solving the eigenvector equation could be inefficient from a 
practical point of view since the computational time to solve 
the eigenvector equation could be too long.
The power iteration scheme that we describe and prove here is 
a practical method to make a fast estimate of the eigenvector, $\bfv_1$ 
associated with the largest eigenvalue, $a_1$ fast without solving 
the eigenvector equation for the case of large $n$.  

Let us start with an initial arbitrary vector, $\bu^{0}$.  
We can construct a new vector, $\bu^{1}$ out of $\bA$ and $\bu^{0}$ 
such that $\bu^{1} = \bA\bu^{0}$.  
Now, using the eigenvectors of $\bA$ as a basis, we can expand $\bu^{0}$ 
such that $\bu^{0} = \sum_{i=1}^{n}b_i\bfv_{i}$. 
So, we can write $\bu^{1}$ such that 
$\bu^{1} = \bA\sum_{i=1}^{n}b_i\bfv_{i} =\sum_{i=1}^{n}a_{i}b_i\bfv_{i}$.
Iterating this process $m$ times leads to a $m$-th vector, $\bu^{m}$ 
such that $\bu^{m} = \sum_{i=1}^{n}a_{i}b_i\bfv_{i}$.  Since $a_1$ is 
the largest eigenvalue, the first component proportional to $a^{m}_{1}$ 
dominates.  Thus, if we iterate this process sufficiently large times,  
$\bu^{m}$ converges effectively to the eigenvector associated with 
the largest eigenvalue.  After $m$ iterations, the fractional error 
caused by approximating $\bfv_1$ by $\bu^{m}$ is proportional to 
$(a_1/a_2)^{m}$, which goes to zero as $m$ becomes large. 

Here, the key assumption made for this power iteration to function  
is that $\bA$ is positive definite, which guarantees $a_i > 0$ for 
all $i=1, n$.  However, even for the case one has a matrix which 
is not positive definite so that not all eigenvalues are positive,  
one can still use the power iteration to find the maximum eigenvector
as far as the largest eigenvalue is positive. It can be made  
by inserting secondary steps between each iteration such that
\beq
\bu^{m+1/2} = \bA\bu^{m}, \hspace{0.5cm}  
\bu^{m+1} = \frac{\bu^{m+1/2}}{|\bu^{m+1/2}|} + \bu^{m-1}, 
\eeq 
with the assumption that not all eigenvalues are negative. 
After the first iteration, we have $\bu^{1} = \sum_{i=1}^{n}
[1 + a_i/(\sum_{j=1}^{n} a^{2}_{j}b^{2}_{j})]b_{i}\bfv_{i}$. 
If $a_i  < 0$, then $1 + a_i/(\sum_{j=1}^{n}a^{2}_{j}b^{2}_{j}) < 1$. 
Thus, this refined power iteration effectively suppresses the 
eigenvectors associated with the negative eigenvalues and 
converges $\bu^{m}$ to $\bfv_{1}$. 

\section{DENSITY RECONSTRUCTION ALGORITHM}

Let $\by \equiv (y_0, y_1, \cdots, y_n)$ is a Gaussian random vector  
of length $n+1$.  The conditional expectation value of the first 
component given the rest components, 
$\langle y_0 | y_1, \cdots, y_n \rangle$ can be calculated such that : 
\ben 
\langle y_0 | y_1, \cdots, y_n \rangle 
&=& \int y_0 P(y_0 | y_1, \cdots, y_n) dy_0,  \nonumber \\
&=& \int y_0 \frac{P(y_0, y_1, \cdots, y_n)}
{P(y_1, \cdots, y_n)} dy_0. \nonumber \\
\label{eqn:cond} 
\een
Here, since $\by$ is Gaussian, 
$\by^{\pr} \equiv (y_1, \cdots, y_n)$ is also Gaussian: 
\beq 
P(y_0, \cdots, y_n) = \frac{1}{\sqrt{(2\pi)^{n+1}|\bV|}}
\exp\left( -\frac{\by^T\cdot\bV^{-1}\cdot\by}{2}\right),
\eeq
\beq 
P(y_1, \cdots, y_n) = \frac{1}{\sqrt{(2\pi)^{n}|\bV^{\pr}|}}
\exp\left( -\frac{\by^{\pr T}\cdot\bV^{\pr -1}\cdot\by^{\pr}}{2}\right),
\eeq
where  
$\bV$ is the $(n+1)\times (n+1)$ covariance matrix for $\by$, and 
$\bV^{\pr}$ is the $n\times n$ covariance matrix for $\by^{\pr}$ 
such that $V_{\mu\nu} = V^{\pr}_{\mu\nu}$ for 
$\mu,\nu = 1, \cdots, n$, and $|\bV|$, $|\bV^{\pr}|$ represent the 
determinants of $\bV$, $\bV^{\pr}$ respectively.
In this Appendix, the Greek indices $\mu,\nu,\tau$ run from $1$ to $n$.  

Let $\bU \equiv \bV^{-1}$.  Then, we can rewrite 
\ben
-\frac{\by^{T}\cdot\bV^{-1}\cdot\by}{2} 
&=& -\frac{\by^{T}\cdot\bU\cdot\by}{2}, \nonumber \\
&=& -\frac{U_{00}y_{0}^2}{2} - U_{0\nu}y_{\nu}y_{0} - 
\frac{U_{\mu\nu}y_{\mu}y_{\nu}}{2}, \nonumber \\  
&=& -\frac{U_{00}}{2}\left( y_0 + \frac{U_{0\nu}y_{\nu}}{U_{00}}\right)^2 
- \frac{1}{2U_{00}}\left( U_{00}U_{\mu\nu} - U_{0\nu}U_{0\nu}\right)
y_{\mu} y_{\nu}, 
\een

But, we have: 
\beq
\frac{1}{U_{00}}\left( U_{00}U_{\mu\nu} - U_{0\mu}U_{0\nu}\right) 
= V_{\mu\nu}^{\pr -1}.
\label{eqn:iden}
\eeq
which can be proved by  
\ben
V^{\pr}_{\mu\nu}\frac{(U_{00}U_{\nu\tau} - U_{0\nu}U_{0\tau})}{U_{00}}
&=& V^{\pr}_{\mu\nu}U_{\nu\tau}-\frac{V_{\mu\nu}U_{\nu0}U_{0\tau}}{U_{00}}, 
\nonumber \\
&=& \delta_{\mu\tau}-\frac{\delta_{\mu0}U_{0\tau}}{U_{00}}=\delta_{\mu\tau},  
\een
since $\mu \neq 0$, i.e.,  $\delta_{\mu0} = 0$. 

Therefore, we can express the conditional probability density 
distribution, \\ $P(y_0 | y_1, \cdots, y_n)$ such that 
\beq 
P(y_0|y_1,\cdots, y_n) 
= \frac{1}{\sqrt{2\pi |\bV||\bV^{\pr}|^{-1}}}
\exp\bigg{\{} -\frac{U_{00}}{2}\left(y_0 + \frac{U_{0\nu}y_\nu}{U_{00}}
\right)^2\bigg{\}}.
\eeq
Thus, equation (\ref{eqn:cond}) can be simplified into
\ben
\langle y_0 | y_1, \cdots, y_n \rangle 
&=& \frac{1}{\sqrt{2\pi |\bV||\bV^{\pr}|^{-1}}}
\int_{-\infty}^{\infty}
y_{0}\exp\bigg{\{} -\frac{U_{00}}{2}\left(y_0+\frac{U_{0\nu}y_\nu}{U_{00}}
\right)^2\bigg{\}}dy_{0}, \nonumber \\
&=& -\frac{1}{\sqrt{|\bV||\bV^{\pr}|^{-1}}}
\frac{U_{0\nu}y_\nu}{U_{00}^{3/2}}.
\label{eqn:cew}
\een

To calculate $|\bV||\bV^{\pr}|^{-1}$, let us construct a $(n+1)\times (n+1)$ 
matrix, say $\bV^{\pr\pr}$ from the $n\times n$ maxtrix $\bV^{\pr}$ 
such that 
\[ \bV^{\pr\pr} \equiv \left(\begin{array}{cc}1, & {\bf 0} \\ 
{\bf 0}, & \bV \end{array} \right). \]
Then, obviously $|\bV^{\pr\pr}| = |\bV^{\pr}|$, i.e., 
$|\bV^{\pr\pr}|^{ -1} = |\bV^{\pr}|^{ -1}$. \\ 
Thus one can say 
$|\bV||\bV^{\pr}|^{-1} = |\bV||\bV^{\pr\pr}|^{-1} = 
|\bV||\bV^{\pr\pr -1}| = |\bV\cdot\bV^{\pr\pr -1}|$. 
But, $\bV\cdot\bV^{\pr\pr -1}$ is a $(n+1) \times (n+1)$ matrix such that 
\[ \bV\cdot\bV^{\pr\pr -1} = \left(\begin{array}{cc}V_{00}, & \bV_{0\tau} \\ 
 \bV_{\nu 0}, & \bI \end{array} \right), \]
where $\bI$ is a $n \times n$ identity matrix. 
Its determinant is straightforwardly calculated to be 
\beq 
|\bV\cdot\bV^{\pr\pr -1}| = V_{00} - V_{0\nu}V_{0\tau}V^{\pr -1}_{\nu\tau}.
\label{eqn:cez}
\eeq 
Hence, we finally find equation (\ref{eqn:cew}) equal to 
\beq
\langle y_0 | y_1, \cdots, y_n \rangle = 
-\frac{U_{0\nu}y_{\nu}}{U_{00}}, 
\label{eqn:cey}
\eeq
since through equations (\ref{eqn:iden}) and (\ref{eqn:cez}), 
\ben
U_{00}|\bV||\bV^{\pr}|^{-1} &=& 
U_{00}(V_{00} - V_{0\nu}V_{0\tau}V^{\pr -1}_{\nu\tau}), 
\nonumber \\ &=& U_{00}V_{00} - V_{0\nu}V_{0\tau}
\left( U_{00}U_{\nu\tau} - U_{0\nu}U_{0\tau}\right), \nonumber \\
&=& 1.   
\een
 
We emphasize that this expression for 
$\langle y_0 | y_1, \cdots, y_n \rangle$ given as equation 
(\ref{eqn:cey}) here is equivalent to equation (10) in  
Betschinger (1987). Bertschinger's formula 
is written as\\ $\langle y_0 \vert {\bf y'}\rangle= V_{0\nu} 
V{^{-1}_{\nu\nu'}}y_{\nu'}$. Let us reexpress the matrix ${\bf V}$ 
such that
\[ {\bf V} = \left(\begin{array}{cc}V_{00}, & {\bf V}_{0i} \\ 
{\bf V}_{i0}, & {\bf V}_{ij}\end{array} \right). \]
Then, one can say ${\bf V}^{-1}_{0i}/{\bf V}^{-1}_{00} = 
\vert {\bf A}^{0i} \vert$, where 
${\bf A}$ represents a cofactor of ${\bf V}$. Now the \\determinant 
of  the cofactor can be expressed as a sum such that 
$\vert {\bf B}\vert = \sum_{i}{\bf B}_{0i} \vert {\bf C}^{0i}\vert$ where 
${\bf B}_{ij} \equiv {\bf A}^{0i}$ and ${\bf C}$ is the cofactor of 
${\bf B}$.  This is explicitly the same as Betschinger's formula, when 
we expand each of Betschinger's matrix elements as the determinants 
of cofactors, and sum it over ${\bf V}_{0\nu}$.  

However, there is one advantage of our formula over Bertschinger's: 
Our conditional expectation value is often computationally much cheaper, 
since for a translation invariant random field,  ${\bf V}^{-1}$ 
can be computed with the fast fourier transformation method (FFT), 
while that is not generally possible for Bertschinger's formula. 

\section{SPIN-SPIN CORRELATION}

Let us first consider the three dimensional spatial spin-spin correlation, \\
$\langle |\hbL(\bx)\cdot\hbL(\bx + \br)|^{2}\rangle \equiv   
\langle \hL_{i}\hL_{i}^{\pr}\hL_{j}\hL_{j}^{\pr}\rangle$. 
Before estimating it theoretically, it is instructive to understand 
the interpretation of $\hbL$ in this expression. In practice, $\hbL$ is 
the measured unit spin vector of an observed galaxy. 
In theory, however, there is no way to calculate the unit spin vector 
of an observed galaxy analytically since galaxy formation 
is still an unsolved problem.  
What one can do at most (and at best) theoretically is to use some 
analytic estimation formula for the expectation value of $\hbL$. 
Here we use equation (\ref{eqn:lij}) as an theoretical formula 
for $\hbL$ based on the linear perturbation theory.  

In equation (\ref{eqn:lij}) we have taken the ensemble average over
the inertia tensors to obtain a result that depends on shear tensors.
Taking the ensemble average of  equation (\ref{eqn:lij}) 
over the shear tensors as well would result
in a trivial $\langle\hL_i \hL_j\rangle=\delta_{ij}/3$.
\ben 
\langle\hL_{i}\hL_{i}^{\pr}\hL_{j}\hL_{j}^{\pr}\rangle 
&=& 
\langle\hL_{i}\hL_{j}\hL_{i}^{\pr}\hL_{j}^{\pr}\rangle,
\label{eqn:inert}\\
&=& 
\langle\left(\frac{1+a}{3}\delta_{ij}-
a\hT_{ik} \hT_{kj}\right)\left(\frac{1+a}{3}\delta_{ij}-
a\hT_{il}^{\pr}\hT_{lj}^{\pr}\right)\rangle,
\label{eqn:ttcorr}\\
&=& 
\frac{1}{3} -\frac{a^2}{3} + a^{2}
\langle\hT_{ik}\hT_{kj}\hT_{il}^{\pr}\hT_{lj}^{\pr}\rangle. 
\label{eqn:3dss}
\een 
We took the expectation value over the inertia tensors going from
(\ref{eqn:inert}) to (\ref{eqn:ttcorr}) assuming the inertia tensors 
at two different positions to be independent. 

In equation (\ref{eqn:3dss}) it is formidable to calculate 
$\langle\hT_{ik}\hT_{kj}\hT_{il}^{\pr}\hT_{lj}^{\pr}\rangle$   
analytically since $\hbT$ is in general not a Gaussian 
random field while $\tbT$ is.   
We approximate 
$\langle\hT_{ik}\hT_{kj}\hT_{il}^{\pr}\hT_{lj}^{\pr}\rangle 
= \langle(\tT_{ik}\tT_{kj}\tT_{il}^{\pr}\tT_{lj}^{\pr})/
(|\tbT|^{2}|\tbT^{\pr}|^{2})\rangle$ by 
$\langle\tT_{ik}\tT_{kj}\tT_{il}^{\pr}\tT_{lj}^{\pr}\rangle
/\langle|\tbT|^{2}\rangle^{2}$,   
and apply the Wick theorem such that 
\ben 
\langle\hT_{ik}\hT_{kj}\hT_{il}^{\pr}\hT_{lj}^{\pr}\rangle 
&=& 
\bigg{\langle}\frac{\tT_{ik}\tT_{kj}\tT_{il}^{\pr}\tT_{lj}^{\pr}}
{|\tbT|^{2}|\tbT^{\pr}|^{2}}\bigg{\rangle}\nonumber, \\
&\approx& 
\frac{\langle\tT_{ik}\tT_{kj}\tT_{il}^{\pr}\tT_{lj}^{\pr}\rangle}
{\langle|\tbT|^{2}|\tbT^{\pr}|^{2}\rangle},   
\nonumber \\
&=& 
\frac{9}{4\xi^{2}(0)}\left(\langle\tT_{ik}\tT_{kj}\rangle
\langle\tT_{il}^{\pr}\tT_{lj}^{\pr}\rangle + 
\langle\tT_{ik}\tT_{il}^{\pr}\rangle
\langle\tT_{kj}\tT_{lj}^{\pr}\rangle + 
\langle\tT_{ik}\tT_{lj}^{\pr}\rangle 
\langle\tT_{kj}\tT_{il}^{\pr}\rangle\right),   
\label{eqn:wic2}
\een
since $\langle|\tbT|^{2}\rangle^{2} = 4\xi^{2}(0)/9$  
by  $\langle\tT_{ij}\tT_{kl} \rangle = 
(3\delta_{ik}\delta_{jl} + 3\delta_{il}\delta_{jk} 
- 2\delta_{ij}\delta_{kl})\xi(0)/45$ \cite{bar-etal86}.   

The $r$-independent term in equation (\ref{eqn:wic2}) is 
straightforwardly calculated such that 
\beq  
\frac{9\langle\tT_{ik}\tT_{kj}\rangle
\langle\tT_{il}^{\pr}\tT_{lj}^{\pr}\rangle
}{4\xi^{2}(0) }
= \frac{1}{3}.
\label{eqn:rin}
\eeq       
The rest $r$-dependent two terms in equation (\ref{eqn:wic2}) can be 
calculated using the given $\tC_{ijkl}$ (Appendix B). We find 
\ben
\frac{9}{4\xi^{2}(0)}
\left(\langle\tT_{ik}\tT_{il}^{\pr}\rangle
\langle\tT_{kj}\tT_{lj}^{\pr}\rangle + 
\langle\tT_{ik}\tT_{lj}^{\pr}\rangle 
\langle\tT_{kj}\tT_{il}^{\pr}\rangle\right) &=&
\frac{9}{4\xi^{2}(0)}\left(\tC_{ikil}\tC_{kjlj} + 
\tC_{iklj}\tC_{kjil} \right) \nonumber \\
&=& \frac{9}{4\xi^{2}(0)}\tC_{iklj}\tC_{kjil}, 
\label{eqn:ess}
\een
since $\tC_{ikil} = \tC_{kjlj} = 0$. 
Now using equation (\ref{eqn:scccor}), one can show that 
\ben
\tC_{iklj}\tC_{kjil} &=&  
\frac{4}{9}\xi^{2} + \frac{14}{5}J_{5}^2 - 4J_{3}J_{5} 
 + \frac{14}{9}J_{3}^2 - \frac{8}{5}J_{5}\xi + \frac{8}{9}J_{3}\xi.
\label{eqn:yss}
\een  
For the case of a power law spectrum of $\xi(r) \propto r^{n}$, every 
term in equation (\ref{eqn:yss}) is proportional to $\xi^{2}(r)$ 
since  $J_{3} = \frac{3}{n+3}r^{n} \propto \xi(r)$ and 
$J_{5} = \frac{5}{n+5}r^{n} \propto \xi(r)$. 
Thus, one can say that for the case of a power law spectrum, 
equation (\ref{eqn:ess}) is proportional to $\xi^{2}(r)$. 
Let us say equation (\ref{eqn:ess}) can be written as 
$\frac{1}{3} + \beta\frac{\xi^{2}(r)}{\xi^{2}(0)}$  
where $\beta$ is a proportionality constant to be determined.  
Note that it is true only for the case of a power-law spectrum.   

In order to optimize the approximation used in equation (\ref{eqn:wic2}), 
we determine the optimal value of the proportionality constant 
by considering the real value of 
$\langle\hT_{ik} \hT_{kj}\hT_{il}^{\pr}\hT_{lj}^{\pr}\rangle$ 
in the limit of $r = 0$.  In the asymptotic limit of $r = 0$, we have 
\beq 
\langle\hT_{ik} \hT_{kj}\hT_{il}^{\pr}\hT_{lj}^{\pr}\rangle 
= \langle\hT_{ik} \hT_{kj}\hT_{il}\hT_{lj}\rangle 
= \frac{1}{2}\langle\hT_{lj}\hT_{lj}\rangle = \frac{1}{2}, 
\label{eqn:hlf} 
\eeq
by equation (\ref{eqn:lem1}). But we also have 
\beq 
\langle\hT_{ik} \hT_{kj}\hT_{il}^{\pr}\hT_{lj}^{\pr}\rangle 
\approx \frac{1}{3} + \beta\frac{\xi^{2}(r)}{\xi^{2}(0)}.
\label{eqn:aphlf}
\eeq
By equating equation (\ref{eqn:aphlf}) with $r=0$ to (\ref{eqn:hlf}), 
we find the best-approximation proportionality constant, $\beta = 1/6$. 

So far we have treated a galaxy as a point-like object.
However, a real galaxy has a finite size with a typical 
one dimensional Lagrangian scale of $0.55 h^{-1}$ Mpc.  Dealing with 
a real galaxy with a finite size amounts to replacing  
$\xi$ by $\xi_{R}$ (a top-hat convolved density correlation 
on a galaxy scale of $R$). 
Finally, we find the approximate estimation of 
the three dimensional spatial spin-spin correlation through equations 
(\ref{eqn:3dss}) to (\ref{eqn:ess}) such that 
\ben 
\langle\hL_{i}\hL_{i}^{\pr}\hL_{j}\hL_{j}^{\pr}\rangle 
&=& 
\frac{1}{3} -\frac{a^2}{3} + a^{2}
\langle\hT_{ik}\hT_{kj}\hT_{il}^{\pr}\hT_{lj}^{\pr}\rangle \nonumber \\
&\approx& \frac{1}{3} -\frac{a^2}{3} + a^{2}
\left(\frac{1}{3} + \frac{1}{6}\xi^{2}_{R}(r)\right) \nonumber \\
&\approx& \frac{1}{3} +\frac{a^2}{6}\xi^{2}_{R}(r),    
\label{eqn:fin}
\een 
with the normalization of $\xi_{R}(0) = 1$.
 
\begin{figure}
\plotone{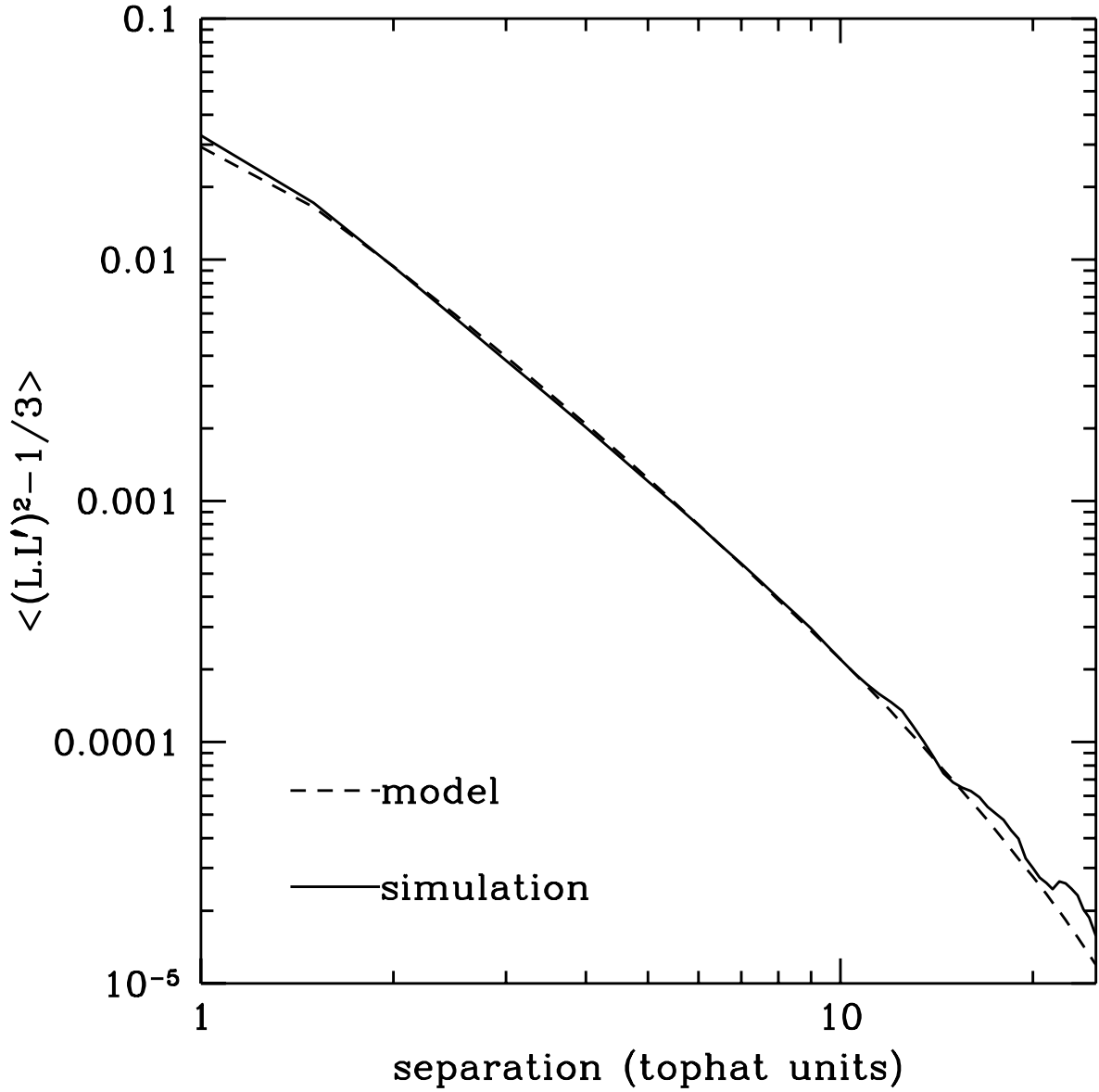}
\caption{The numerical verification of the approximation 
used in equation (\ref{eqn:wic2}). The solid line shows the correlation 
function of spins on a random Gaussian lattice, while the dotted line 
is the theoretical model from equation (\ref{eqn:fin}).\label{fig3}}
\end{figure}

Note that the only approximation made 
in the derivation of equation (\ref{eqn:fin}) is to replace 
$\langle\hT\hT\hT^{\pr}\hT^{\pr}\rangle$ with 
$\langle\tT\tT\tT^{\pr}\tT^{\pr}\rangle/\langle|\tT|^2\rangle^2$.
We have tested the validity of this approximation by Monte-Carlo
simulation, and found that it holds good for all value of $r$.  We
generated a three dimensional Gaussian shear field with the density 
correlation function $\xi (r) \propto r^{-1}$, and smoothed it 
with a top-hat window on a scale radius of two grids.   
We then generate the unit spins by multiplying the shear tensors 
with the uncorrelated inertia tensors with an approximate normalization.   
We then plot the LHS of equation (\ref{eqn:fin}) using the randomly
generated random fields shown as the solid line in Fig. 3, 
while the RHS derived from the density correlation function is shown 
as the dotted line.  In this model we have used $a=3/5$. 
We note the excellent agreement between the model and the simulation 
results, showing that the application of Wick's theorem to the unit shears 
in equation (\ref{eqn:wic2}) is well justified.

With the similar manner one can also find a two dimensional 
spin-spin correlation, $\langle |\hbS(\bx)\cdot\hbS(\bx + \br)|^{2}\rangle$. 
For the case of a power law spectrum,  
all quadratic statistics in the spin should depend on the square of 
the density correlation function.
It is reasonable to assume that the two dimensional spin-spin correlation 
should be also expressed as $1/2 + A\xi_{R}^{2}(r)$ for powerlaw
correlations.  Here $1/3$ is replaced by $1/2$ since  
$\langle |\hbS(\bx)\cdot\hbS(\bx + \br)|^{2}\rangle = 1/2$ 
for the two dimensional random spins. 

Through equations (\ref{eqn:2d1ij}) and (\ref{eqn:2d2ij}), 
we have $\langle \hS^2_{1}-\hS^2_{2}|\hbT\rangle 
=  - 5a(\hT_{1k}\hT_{k1} -\hT_{2k}\hT_{k2} )/4$, 
$ \langle\hS_{1}\hS_{2}|\hbT\rangle = 
- 5a\hT_{1k}\hT_{k2}/4$.   
The comparison of this two dimensional expression with 
equation (\ref{eqn:lij}) indicates that the shear-spin correlation 
parameter $a$ should be replaced by $5a/4$ for the two dimensional case.  
In other words, the amplitude of $A = a^2/6$ 
for the three dimensional case should be replaced by $A = (5a/4)^2/6$ 
for the two dimensional case.  
Thus, we have 
\beq 
\langle \hS_{i}\hS_{j}\hS_{i}^{\pr}\hS_{j}^{\pr}\rangle 
 \approx \frac{1}{2} + \frac{25}{96}a^2\xi_{R}^{2}(r).
\eeq

\section{SPIN-DIRECTION CORRELATION}

The same technique used in Appendix H can be also applied 
to calculate the spin-direction correlation, 
$\langle \hL_{i}\hL_{j}\hr_{i}\hr_{j}\rangle$ such that 
\ben 
\langle \hL_{i}\hL_{j}\hr_{i}\hr_{j}\rangle 
&=& 
\langle\left(\frac{1+a}{3}\delta_{ij} - a\hT_{ik}^{R}\hT_{kj}^{R}\right)
\left(\frac{1-b}{3}\delta_{ij} + b\hT_{ik}^{R^{\pr}}\hT_{kj}^{R^{\pr}}\right)
\rangle, \\
&=& 
\frac{1}{3} + \frac{ab}{3} -    ab\langle\hT_{ik}^{R}\hT_{kj}^{R}
\hT_{il}^{R^{\pr}}\hT_{lj}^{R^{\pr}}\rangle.    
\een     

Using the same approximation that is used in equation (\ref{eqn:wic2}) 
and verified by Monte-Carlo simulations (see Fig. 3), we have 
\ben 
\frac{9\langle\hT_{ik}^{R}\hT_{kj}^{R}\hT_{il}^{R^{\pr}}
\hT_{lj}^{R^{\pr}}\rangle}
{4(\sigma_{R}\sigma_{R^{\pr}})^{2}}  
&=&
\frac{9\langle\tT_{ik}^{R}\tT_{kj}^{R}\tT_{il}^{R^{\pr}}
\tT_{lj}^{R^{\pr}}\rangle}
{4(\sigma_{R}\sigma_{R^{\pr}})^{2}},  
\nonumber \\
&=& 
\frac{9(\langle\tT_{ik}^{R}\tT_{kj}^{R}\rangle
\langle\tT_{il}^{R^{\pr}}\tT_{lj}^{R^{\pr}}\rangle +
\langle\tT_{ik}^{R}\tT_{il}^{R^{\pr}}\rangle
\langle\tT_{kj}^{R}\tT_{lj}^{R^{\pr}}\rangle + 
\langle\tT_{ik}^{R}\tT_{lj}^{R^{\pr}}\rangle 
\langle\tT_{kj}^{R}\tT_{il}^{R^{\pr}}\rangle)} 
{4(\sigma_{R}\sigma_{R^{\pr}})^{2}}.  
\label{eqn:wicg2}
\een 
Note that we calculate the shear correlation at the same  position  
but smoothed on two different scales.  
Following the same logic explained in Appendix H,  we have \\
$9\langle\tT_{ik}^{R}\tT_{kj}^{R}\rangle
\langle\tT_{il}^{R^{\pr}}\tT_{lj}^{R^{\pr}}
\rangle/(4\sigma_{R}^{2}\sigma_{R^{\pr}}^{2}) 
= 1/3$, \\ 
$9(\langle\tT_{ik}^{R}\tT_{il}^{R^{\pr}}\rangle
\langle\tT_{kj}^{R}\tT_{lj}^{R^{\pr}}\rangle + 
\langle\tT_{ik}^{R}\tT_{lj}^{R^{\pr}}\rangle
\langle\tT_{kj}^{R}\tT_{il}^{R^{\pr}}\rangle)
/(4\sigma_{R}^{2}\sigma_{R^{\pr}}^{2})  
\propto \langle\delta^{R}(\bx)\delta^{R^{\pr}}(\bx)\rangle^{2}/
(\sigma_{R}^{2}\sigma_{R^{\pr}}^{2})$.   
The proportionality constant can be easily obtained to be $1/6$ 
again by considering the limit of $r=0$ (i.e., $R^{\pr} = R$).
 
So, finally we find 
\beq
\langle \hL_{i}\hL_{j}\hr_{i}\hr_{j}\rangle =  \frac{1}{3} + 
\frac{ab}{6}\frac{\langle\delta_{R^{\pr}}\delta_{R}\rangle^{2}}
{(\sigma_{R^{\pr}}\sigma_{R})^2}.
\label{eqn:llrr}
\eeq
For the two dimensional case, 
replacing $1/3$ with $1/2$, and $a$ with $5a/4$, we also find 
\beq
\langle \hS_{i}\hS_{j}\hr_{i}\hr_{j}\rangle =   \frac{1}{2} + 
\frac{5ab}{24}\frac{\langle\delta_{R^{\pr}}\delta_{R}\rangle^{2}}
{(\sigma_{R^{\pr}}\sigma_{R})^2}.
\eeq

\section{CORRELATION PARAMETERS}

In this final appendix, we provide the derivations of the optimal 
estimation formula (eq. [\ref{eqn:opa}] and eq. [\ref{eqn:opb}]) for 
the two correlation parameters ($a$ and $b$ respectively), and the 
involved error-bar formula as well. 

Multiplying $\hlam_i$ to each side of equation (\ref{eqn:lij}), and using 
$\sum_{i}{\hlam_{i}}=0$, $\sum_{i}{\hlam^{2}_{i}}=1$, 
$\sum_{i}{\hlam^{4}_{i}}=1/2$, $\sum_{i < j}{\hlam_{i}\hlam_{j}}=-1/2$, 
and $\sum_{i < j}{\hlam^{2}_{i}\hlam^{2}_{j}}=1/4$, 
equation (\ref{eqn:opa}) is straightforwardly derived 
from equation (\ref{eqn:lij}). Here note that equation (\ref{eqn:lij}) 
is used as a theoretical estimation formula for $\hL_{i}\hL_{j}$.
The error $\epsilon_{a}$ involved in the measurement of the average 
value of $a$ is given as the standard deviation of $a$ for the case of 
no correlation between $\hbT$ and $\hbL$.  

For the case of no correlation, $\langle a \rangle = 0$. 
So, $\epsilon_{a} = \sqrt{\langle a^2 \rangle}$. 
Now, by equation (\ref{eqn:opa}) we have  
\beq 
\epsilon_{a} = \sqrt{\langle a^2 \rangle} 
= \sqrt{\langle (2-6\sum_{i}\hlam_{i}^{2}\hL_{i}^{2})^{2}\rangle} 
= \sqrt{\frac{4}{5}},  
\eeq  
since $\langle\hL_{i}^{2}\rangle = 1/3$, 
$\langle\hL_{i}^{4}\rangle = 1/5$ for each $i = 1,2,3$ and 
$\langle\hL_{i}^{2}\hL_{j}^{2}\rangle = 1/15$ for each $i \ne j$   
if $\hbL$ is random. Thus, for the $N_t$ ensemble, 
we finally have $\epsilon_{a} = \sqrt{4/(5N_{t})}$. 

One can derive an optimal formula for $b$ with the similar argument. 
But, there is one notable difference in deriving the optimal 
formula for $b$. For $a$, we have directly used equation  (\ref{eqn:lij}) 
to derive equation (\ref{eqn:opa}) where each $|\hL_{i}|^2$ is 
weighed by the square of the eigenvalue, $\hlam_{i}^{2}$. 
It is adequate since the unit galaxy spin vector, $\hbL$, is expected  
to be aligned with the intermediate principal axis of the shear tensor 
which does not involve any degeneracy upon a sign change of the shear 
tensor ($\hbT \rightarrow -\hbT$).   
Whereas, the unit galaxy separation vector, $\hbr$, is expected 
to align with the major principal axis of the  shear tensor, 
and thus orthogonal to the minor principal axis.   
But the major and minor principal axes of the shear tensor are 
interchanged by a sign change of the shear tensor.  

In order to measure the real correlation between 
$\hbr$  and  the major principal axis of the shear tensor, 
we first define a secondary correlation parameter, $d$, by weighing
each  $\hr_{i}$ by $\hlam_{i}$ instead of $\hlam_{i}^2$ such that 
\beq
d = \hlam_{i}\hr_{i}^{2}.   
\label{eqn:opbj}
\eeq
In the following we show that in fact $b = d/\sqrt{2}$, and 
prove that the optimal estimation formula for $b$ is indeed 
equation (\ref{eqn:opb}). 

Let $\be^{M}$ be the major eigenvector of $\hbT^{R^{\pr}}$. 
Since $\hbr$ is supposed to be aligned with $\be^{M}$, one can model 
the unit galaxy separation vector as a mixture of the major 
eigenvector with a random component such that   
\beq
\langle \hr_i \hr_j | \hbT^{R^{\pr}} \rangle = 
\gamma^2 e^{M}_i e^{M}_j + (1-\gamma^2)\frac{\delta_{ij}}{3}
\label{eqn:crij}
\eeq
where $\gamma^2$ measures the strength of the correlation between 
$\hbr$ and $\be^{M}$.  Using $\be^{M}=(1,0,0)$ in the principal axis 
frame of $\hbT^{R^{\pr}}$, one gets 
$\langle \hr_{1}^{2} | \hbT^{R^{\pr}} \rangle = 
\frac{1}{3} + \frac{2\gamma^2}{3}$, 
$\langle \hr_{2}^{2} | \hbT^{R^{\pr}} \rangle = 
\frac{1}{3} - \frac{\gamma^2}{3}$,  
and $\langle \hr_{2}^{2} | \hbT^{R^{\pr}} \rangle = 
\frac{1}{3} - \frac{\gamma^2}{3}$. 
Now, using equation (\ref{eqn:crij}) for the theoretical estimation  
of $|\hr_{i}|^{2}$ in equation  (\ref{eqn:opbj}), we have 
\beq
d = \gamma^2\left( \frac{2}{3}\hlam_{1} - \frac{1}{3}
\hlam_{2} - \frac{1}{3} \hlam_{3}\right).
\label{eqn:opd}
\eeq
Approximating the three eigenvalues of the traceless shear tensor
as $\hlam_1=1/\sqrt{2} = -\hlam_3$ and $\hlam_3 = 0$ since 
$\langle \hlam_1^2\rangle = \langle \hlam_{3}^2 \rangle \gg 
\langle \hlam_{2}^2\rangle$ and $\hlam_{i}\hlam_{i} = 1$, and 
inserting these values of the three eigenvalues into equation 
(\ref{eqn:opd}), we find $\gamma^2=d\sqrt{2}$. 

Now, let us model the unit spin vector as a mixture of the intermediate 
eigenvector of $\hbT^{R}$, $\be^{I}$, with a random component such that
\beq
\langle \hL_i \hL_j | \hbT^{R} \rangle = - \alpha^2 e^{I}_i e^{I}_j + 
(1 + \alpha^2)\frac{\delta_{ij}}{3}, 
\label{eqn:lijn}
\eeq
where $\alpha^{2}$ measures the correlation strength between 
$\hbL$ and $\be^{I}$. Using $\be^{I}=(0,1,0)$ in the principal axis 
frame of $\hbT^{R}$, one gets 
$\langle \hL_{1}^{2} | \hbT^{R} \rangle = \frac{1}{3} + \frac{\alpha^2}{3}$, 
$\langle \hL_{2}^{2} | \hbT^{R} \rangle = \frac{1}{3} - \frac{2\alpha^2}{3}$, 
and $\langle \hL_{2}^{2} | \hbT^{R} \rangle = 
\frac{1}{3} + \frac{\alpha^2}{3}$.
Inserting the above approximate values of the eigenvalues into 
equation (\ref{eqn:lijn}), we have $\alpha^{2} = -a/2$.

Extrapolating equation (\ref{eqn:crij}) to the limit of $R^{\pr} = R$, 
let us calculate 
$\langle \hL_i \hL_j | \hbT^{R} \rangle 
\langle \hr_i \hr_j | \hbT^{R{\pr}} \rangle$ 
at this asymptotic limit of $R^{\pr} = R$.
Using equations (\ref{eqn:llrr}), (\ref{eqn:crij}) and (\ref{eqn:lijn}) 
for the theoretical estimation formula, and given $\be^{M}\cdot\be^{I} = 0$ 
due to the orthogonality of the eigenvectors, 
we have
\beq 
\langle \hL_i \hL_j | \hbT^{R} \rangle
\langle \hr_i \hr_j | \hbT^{R^{\pr}} \rangle = 
\frac{1}{3} - \frac{\alpha^{2}\gamma^{2}}{3} = \frac{1}{3} + \frac{ab}{6}. 
\eeq
It shows that $b = d\sqrt{2}$ since $\alpha^{2} = - \frac{a}{2}$ and 
$\gamma^{2} = \sqrt{2}d$. 

The error $\epsilon_{b}$ involved in the measurement of 
the average of $b$ is also given as the standard deviation of $b$ 
for the case of no correlation between  $\hbT$ and $\hbr$. 
Now we have
\beq
\epsilon_{b} = \sqrt{\langle b^2 \rangle} 
= \sqrt{2\sum_{i,j}\langle\hlam_{i}\hlam_{j}\rangle
\langle\hr_{i}^{2}\hr_{j}^{2}\rangle }
= \sqrt{\frac{4}{15}}.
\eeq  
Thus, for the $N_{t}$ ensemble, $\epsilon_{b} = \sqrt{4/(15 N_{t})}$.

\end{document}